\documentclass[12pt,journal,draftcls,a4paper,onecolumn]{IEEEtran}

\usepackage[all,poly]{xy}
\usepackage{amsmath}
\usepackage{subfigure}
\usepackage{algorithmic}
\usepackage{algorithm}
\usepackage{multirow}
\usepackage{amsfonts}
\usepackage{graphicx}
\usepackage{amssymb}
\usepackage[noadjust]{cite}
\usepackage{float}
\usepackage{color}
\usepackage{array}
\usepackage[stable]{footmisc}
\usepackage{rotate}
\newcommand{\figwidth}{\columnwidth}
\newcommand{\figwidthsmall}{0.85\columnwidth}
\newcommand{\figwidthbis}{\textwidth}

\usepackage{float}
\newcommand{\Y}{\boldsymbol{y}}
\newcommand{\YMAT}{\boldsymbol{Y}}

\newcommand{\SIGN}{\boldsymbol{m}}
\newcommand{\SIGNMAT}{\boldsymbol{M}}

\newcommand{\vnoise}{s}

\newcommand{\ABUND}{a}
\newcommand{\ABUNDVEC}{\boldsymbol{a}}
\newcommand{\ABUNDMAT}{\boldsymbol{A}}

\newcommand{\COEFF}{t}
\newcommand{\COEFFVEC}{\boldsymbol{t}}
\newcommand{\COEFFMAT}{\boldsymbol{T}}

\newcommand{\MCOEFF}{\psi}
\newcommand{\MCOEFFvec}{\boldsymbol{\psi}}
\newcommand{\GMCOEFF}{\boldsymbol{\Psi}}
\newcommand{\VCOEFFvec}{\boldsymbol{\Sigma}}
\newcommand{\VCOEFF}{\sigma^2}

\newcommand{\GVCOEFF}{\boldsymbol{\Sigma}}

\newcommand{\HyperMCOEFF}{\upsilon^2}

\newcommand{\HyperVCOEFFa}{\xi}
\newcommand{\HyperVCOEFFb}{\gamma}
\newcommand{\HyperVAR}{\delta}
\newcommand{\HyperMAT}{\Omega}

\newcommand{\LABEL}{z}
\newcommand{\LABELvec}{\boldsymbol{z}}
\newcommand{\GRAN}{\beta}
\newcommand{\vrnwalk}{u}

\newcommand{\Ik}{\mathcal{I}_k}
\newcommand{\BTHETA}{\boldsymbol{\Theta}}

\newcommand{\transp}{^T}
\newcommand{\norm}[1]{\left\|#1\right\|}

\title{Enhancing hyperspectral image unmixing \\with spatial correlations}

\author{Olivier Eches, Nicolas Dobigeon and
Jean-Yves Tourneret\\
\normalsize University of Toulouse, IRIT/INP-ENSEEIHT/T\'eSA\\
2 rue Camichel, 31071 Toulouse, France\\
\small\texttt{\{Olivier.Eches,Nicolas.Dobigeon,Jean-Yves.Tourneret\}@enseeiht.fr}
\footnote{\copyright 2011 IEEE. Personal use of this material is
permitted. Permission from IEEE must be obtained for all other uses,
in any current or future media, including reprinting/republishing
this material for advertising or promotional purposes, creating new
collective works, for resale or redistribution to servers or lists,
or reuse of any copyrighted component of this work in other works}}

\begin{document}
\maketitle

\begin{abstract}
This paper describes a new algorithm for hyperspectral image
unmixing. Most of the unmixing algorithms proposed in the literature
do not take into account the possible spatial correlations between
the pixels. In this work, a Bayesian model is introduced to exploit
these correlations. The image to be unmixed is assumed to be
partitioned into regions (or \emph{classes}) where the statistical
properties of the abundance coefficients are homogeneous. A Markov
random field is then proposed to model the spatial dependency of the
pixels within any class. Conditionally upon a given class, each
pixel is modeled by using the classical linear mixing model with
additive white Gaussian noise. This strategy is investigated the
well known linear mixing model. For this model, the posterior
distributions of the unknown parameters and hyperparameters allow
ones to infer the parameters of interest. These parameters include
the abundances for each pixel, the means and variances of the
abundances for each class, as well as a classification map
indicating the classes of all pixels in the image. To overcome the
complexity of the posterior distribution of interest, we consider
Markov chain Monte Carlo methods that generate samples distributed
according to the posterior of interest. The generated samples are
then used for parameter and hyperparameter estimation. The accuracy
of the proposed algorithms is illustrated on synthetic and real
data.
\end{abstract}

\begin{IEEEkeywords}
Bayesian inference, Monte Carlo methods, spectral unmixing,
hyperspectral images, Markov random fields, Potts-Markov model.
\end{IEEEkeywords}

\newpage
\section{Introduction}
\label{sec:introduction} Since the early $90$'s, hyperspectral
imagery has been receiving growing interests in various fields of
applications. For example, hyperspectral images have been recently
used successfully for mapping the timber species in tropical
forestry \cite{Jusoff2009}. Hyperspectral image analysis involves
many technical issues such as image classification, image
segmentation, target detection and the crucial step of spectral
unmixing. The problem of spectral unmixing has been investigated for
several decades in both the signal processing and geoscience
communities where many solutions have been proposed (see for
instance \cite{Keshava2002} and \cite{Chang2007} and references
therein). Hyperspectral unmixing consists of decomposing the
measured pixel reflectances into mixtures of pure spectra whose
fractions are referred to as abundances. Assuming the image pixels
are linear combinations of pure materials is very common in the
unmixing framework. More precisely, the linear mixing model (LMM)
considers the spectrum of a mixed pixel as a linear combination of
endmembers \cite{Keshava2002}. The LMM requires to have known
endmember signatures. These signatures can be obtained from a
spectral library or by using an endmember extraction algorithm
(EEA). Some standard EEAs are reviewed in \cite{Martinez2006}. Once
the endmembers that appear in a given image have been identified,
the corresponding abundances have to be estimated in a so-called
\emph{inversion} step. Due to obvious physical considerations, the
abundances have to satisfy positivity and sum-to-one constraints. A
lot of inversion algorithms respecting these constraints have been
proposed in the literature. The fully constrained least squares
(FCLS) \cite{Heinz2001} and scaled gradient (SGA) \cite{Theys2009}
algorithms are two optimization techniques that ensure the
positivity and sum-to-one constraints inherent to the unmixing
problem. Another interesting approach introduced in
\cite{Dobigeon_IEEE_TSP_2008} consists of assigning appropriate
prior distributions to the abundances and to solve the unmixing
problem within a Bayesian framework. However, all these inversion
strategies have been developed in a pixel-by-pixel context and,
consequently, do not exploit the possible spatial correlations
between the different pixels of the hyperspectral image. In this
paper, we show that taking these spatial correlations into account
allows one to improve the unmixing procedure. More precisely, the
Bayesian algorithm initially developed in
\cite{Dobigeon_IEEE_TSP_2008} is modified to introduce spatial
constraints between the abundance coefficients to be estimated.

Within a Bayesian estimation framework, a very popular strategy for
modeling spatial information in an image is based on Markov random
fields (MRFs). MRFs have been widely used in the image processing
literature to properly describe neighborhood dependance between
image pixels. MRFs and their pseudo-likelihood approximations have
been introduced by Besag in \cite{Besag1974}. They have then been
popularized by Geman in \cite{Geman1984} by exploiting the Gibbs
distribution inherent to MRFs. There are mainly two approaches that
can be investigated to model spatial correlations between the
abundances of an hyperspectral image with MRFs. The first idea is to
define appropriate prior distributions for the abundances
highlighting spatial correlations.
This approach has been for instance
adopted by Kent and Mardia in \cite{Kent1988} where several
techniques have been introduced for mixed-pixel classification of
remote sensing data. These techniques rely on a fuzzy membership
process, which implicitly casts the achieved classification task as
a standard unmixing problem\footnote{Note that, to our knowledge,
the Kent and Mardia's paper is one of the earliest work explicitly
dealing with linear unmixing of remotely sensed images.}. Modeling
the abundance dependencies with MRFs makes this approach
particularly well adapted to unmix images with smooth abundance
transition throughout the scene.


Conversely, this paper proposes to
exploit the pixel correlations in an underlying membership model.
This standard alternative strategy allows more flexibility and
appears more suited for images composed of distinct areas, as
frequently encountered in remote sensing applications. Moreover,
this approach has the great advantage of easily generalizing the
Bayesian algorithms previously introduced in
\cite{Dobigeon_IEEE_TSP_2008,Dobigeon_IEEE_TSP_2009}, as detailed
further in the manuscript. It consists of introducing labels that
are assigned to the pixels of the image. Then MRFs are not assigned
on the abundances directly but on these hidden variables, leading to
a softer classification. More precisely, to take into account the
possible spatial correlations between the observed pixels, a
Potts-Markov field \cite{Wu1982} is chosen as prior for the
labels. This distribution enforces the neighboring pixels to belong
to the same class. Potts-Markov models have been extensively used
for classification/segmentation of hyperspectral data in the remote
sensing and image processing literatures
\cite{Mohammadpour2004,Rellier2004,Neher2005,Feron2005,Bali2008,Li2010}.
Note that other research works, such as \cite{Fauvel2008} and
\cite{Tarabalka2010}, have proposed alternative strategies of
modeling spatial correlations between pixels for classification of
hyperspectral images. All these works have shown that taking into
account the spatial correlations is of real interest when analyzing
hyperspectral images.

This paper proposes to study the interest of using MRFs for unmixing
hyperspectral images. More precisely, the Bayesian unmixing strategy
developed in \cite{Dobigeon_IEEE_TSP_2008} is generalized to take
into account spatial correlations between the pixels of a
hyperspectral image. The hyperspectral image to be analyzed is
assumed to be partitioned into homogeneous regions (or classes) in
which the abundance vectors have the same first and second order
statistical moments (means and covariances). This assumption implies
an implicit image classification, modeled by hidden labels whose
spatial dependencies follow a Potts-Markov field. Conditionally upon
these labels, the abundance vectors are assigned appropriate prior
distributions with unknown means and variances that depend on the
pixel class. These prior distributions ensure the positivity and
sum-to-one constraints of the abundance coefficients. They are based
on a reparametrization of the abundance vectors and are much more
flexible than the priors previously studied in
\cite{Dobigeon_IEEE_TSP_2008}, \cite{Eches_IEEE_WHISPERS_2009} or
\cite{Dobigeon_IEEE_TSP_2009}. Of course, the accuracy of the
abundance estimation procedure drastically depends on the
hyperparameters associated to these priors. This paper proposes to
estimate these hyperparameters in a fully unsupervised manner by
introducing a second level of hierarchy in the Bayesian inference.
Non-informative prior distributions are assigned to the
hyperparameters. The unknown parameters (labels and abundance
vectors) and hyperparameters (prior abundance mean and variance for
each class) are then inferred from their joint posterior
distribution. Since this posterior is too complex to derive
closed-form expressions for the classical Bayesian estimators,
Markov chain Monte Carlo (MCMC) techniques are studied to alleviate
the numerical problems related to the LMM with spatial constraints.
MCMC allow one to generate samples asymptotically distributed
according to the joint posterior of interest. These samples are then
used to approximate the Bayesian estimators, such as the minimum
mean square error (MMSE) or the maximum \emph{a posteriori}
estimators. Note that the underlying classification and abundance
estimation problems are jointly solved within this Bayesian
framework.

The paper is organized as follows. The unmixing problem associated
to the LMM with spatial correlations is formulated in
\ref{sec:formulation}. Section~\ref{sec:Bayesian_model} introduces a
hierarchical Bayesian model appropriate to this unmixing problem.
The MCMC algorithm required to approximate the Bayesian LMM
estimators is described in Section \ref{sec:hybrid_gibbs_sampler}.
Simulation results conducted on simulated and real data are provided
in Sections~\ref{sec:simus_synthetic} and \ref{sec:simus_real}.
Finally, conclusions related to this work are reported in
Section~\ref{sec:conclusions}.

\section{Technical background and problem formulation}\label{sec:formulation}

\subsection{Unmixing statistical model}\label{subsec:unmix_models}

As highlighted in the previous section,
the LMM has been mainly proposed in the remote sensing literature
for spectral unmixing. The LMM assumes that the spectrum of a given
pixel is a linear combination of deterministic endmembers corrupted
by an additive noise \cite{Keshava2002} considered here as white
Gaussian. More specifically, the observed $L$-spectrum of a given
pixel $p$ is defined as
\begin{equation}
 \Y_p = \SIGNMAT \ABUNDVEC_p + \boldsymbol{n}_p
\end{equation}
where $L$ is the number of spectral bands, $\SIGNMAT =
[\SIGN_1,\ldots,\SIGN_R]$ is a known $L\times R$ matrix containing
the $L$-spectra of the endmembers, $\ABUNDVEC_p $ is the $R\times 1$
abundance vector,  $R$ is the number of endmembers that are present
in the image and $\boldsymbol{n}_p$ is the noise vector. The vector
$\boldsymbol{n}_p$ is classically assumed to be an independent and
identically distributed (i.i.d.) zero-mean Gaussian sequence with
unknown variance $\vnoise^2$
\begin{equation}
  \boldsymbol{n}_p |\vnoise^2 \sim
\mathcal{N}\left(\boldsymbol{0}_L,\vnoise^2 \boldsymbol{I}_L
\right)
\end{equation}
where $\boldsymbol{I}_L$ is the $L \times L$ identity matrix.
Note that the noise is the same for
all pixels of the hyperspectral image and does not vary from one
pixel to another, which has been a common assumption widely admitted
in the hyperspectral literature
\cite{Harsanyi1994,Chang1998,Chang1998b}.

Considering an image of $P$ pixels, standard matrix notations can be
adopted leading to $\YMAT = [\Y_1,\ldots \Y_P]$ and $\ABUNDMAT =
[\ABUNDVEC_1, \ldots, \ABUNDVEC_P]$.

\subsection{Introducing spatial dependencies between abundances}
\label{subsec:spatial_dependencies} We propose in this paper to
exploit some spatial correlations between the pixels of the
hyperspectral image to be analyzed. More precisely, it is
interesting to consider that the abundances of a given pixel are
similar to the abundances of its neighboring pixels. Formally, the
hyperspectral image is assumed to be partitioned into $K$ regions or
\emph{classes}. Let $\Ik \subset \left\{1,\ldots,P\right\}$ denote
the subset of pixel indexes belonging to the $k$th class. A label
vector of size $P \times 1$ denoted as $\LABELvec =
\left[\LABEL_1,\ldots,\LABEL_P \right]^T$ with $\LABEL_p \in
\left\{1,\ldots,K \right\}$ is introduced to identify the class to
which each pixel $p$ belongs ($p=1,\ldots,P$). In other terms
$\LABEL_p = k$ if and only if $p \in \Ik$. In each class, the
abundance vectors to be estimated are assumed to share the same
first and second order statistical moments, i.e., $\forall
k\in\left\{1,\ldots,K\right\},\ \forall (p,p') \in \Ik\times \Ik$
\begin{equation}
\label{eq:class_statistics}
\begin{split}
   \mathrm{E}\left[\ABUNDVEC_p\right] &=
   \mathrm{E}\left[\ABUNDVEC_{p'}\right] = \boldsymbol{\mu}_k\\
   \mathrm{E}\left[\left(\ABUNDVEC_p-\boldsymbol{\mu}_k\right)\left(\ABUNDVEC_p-\boldsymbol{\mu}_k\right)\transp\right]
   &=
\mathrm{E}\left[\left(\ABUNDVEC_{p'}-\boldsymbol{\mu}_k\right)\left(\ABUNDVEC_{p'}-\boldsymbol{\mu}_k\right)\transp\right].
\end{split}
\end{equation}
Therefore, the $k$th class of the hyperspectral image to be unmixed
is fully characterized by its abundance mean vector and the
abundance covariance matrix.

\subsection{Markov random fields}
\label{subsec:MRF} 
\begin{figure}[htbp]
\centering
\includegraphics[width=\figwidthsmall]{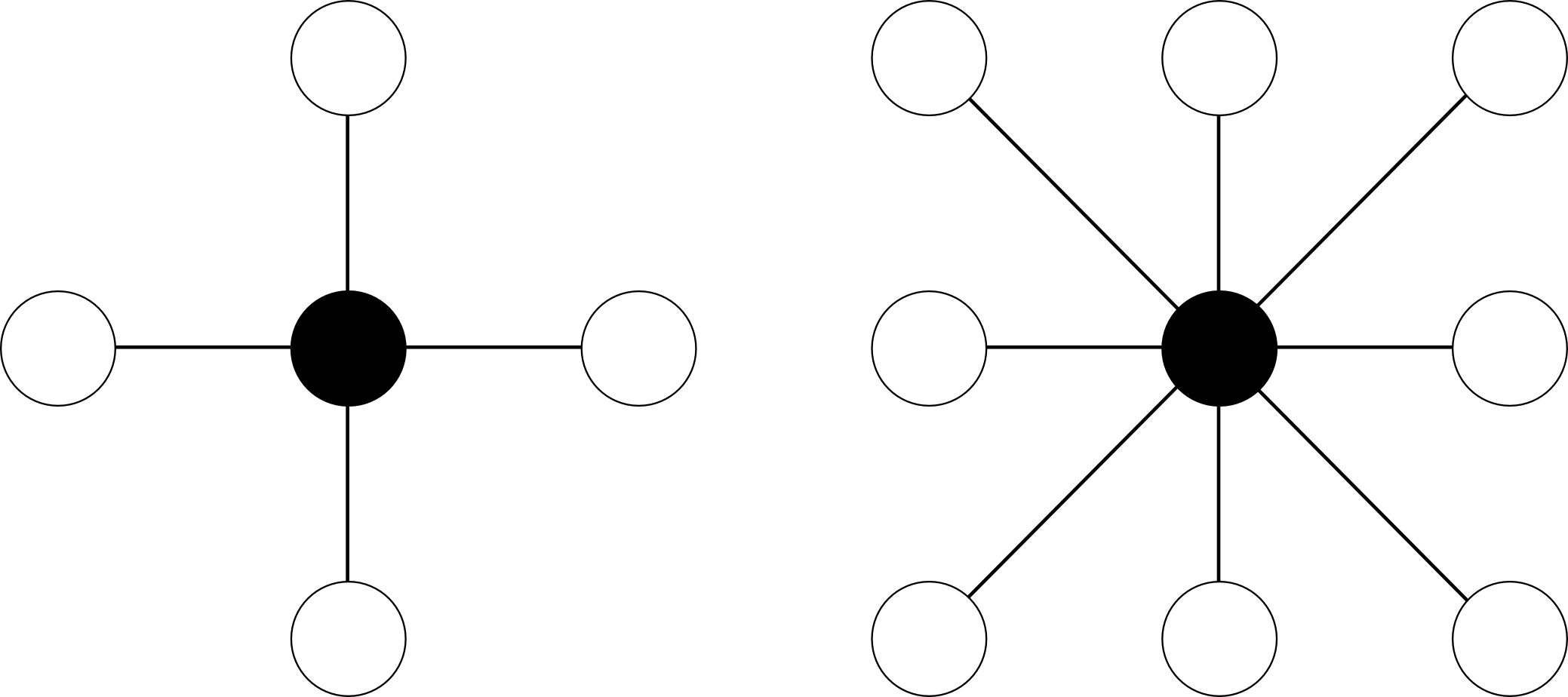} \caption{4-pixel (left) and 8-pixel (right) neighborhood structures. The considered pixel appear as a black circle
 whereas its neighbors are depicted in white.}\label{fig:clique}
\end{figure}

To describe spatial correlations between pixels,
it is important to properly define a neighborhood structure. The
neighborhood relation between two pixels $i$ and $j$ has to be
symmetric: if $i$ is a neighbor of $j$ then $j$ is a neighbor of
$i$. This relation is applied to the nearest neighbors of the
considered pixel, for example the fourth, eighth or twelfth nearest
pixels. Fig. \ref{fig:clique} shows two examples of neighborhood
structures. The four pixel structure or \emph{$1$-order
neighborhood} will be considered in the rest of the paper.
Therefore, the associated set of neighbors, or \emph{cliques}, has
only vertical and horizontal possible configurations (see
\cite{Besag1974,Geman1984} for more details).

Once the neighborhood structure has been established, the MRF can be
defined. Let $z_p$ denote a random variable associated to the $p$th
pixel of an image of $P$ pixels. In the context of hyperspectral
image unmixing, the variables $z_1,\ldots,z_P$ indicate the pixel
classes and take their values in a finite set
$\left\{1,\ldots,K\right\}$ where $K$ is the number of possible
classes. The whole set of random variables
$\left\{z_1,\ldots,z_P\right\}$ forms a random field. An MRF is then
defined when the conditional distribution of $z_i$ given the other
pixels $\boldsymbol{z}_{\textrm{-}i}$ only depend on its neighbors
$\boldsymbol{z}_{\mathcal{V}(i)}$, i.e.,
\begin{equation}
f\left(\LABEL_i | \LABELvec_{\textrm{-}i}\right) = f\left(\LABEL_i |
\LABELvec_{\mathcal{V}(i)}\right)
\end{equation}
where $\mathcal{V}(i)$ is the neighborhood structure considered and
$\LABELvec_{\textrm{-}i} = \{\LABEL_j ; j \neq i \}$.

\begin{figure}[htbp]
\centering
\includegraphics[width=\figwidth]{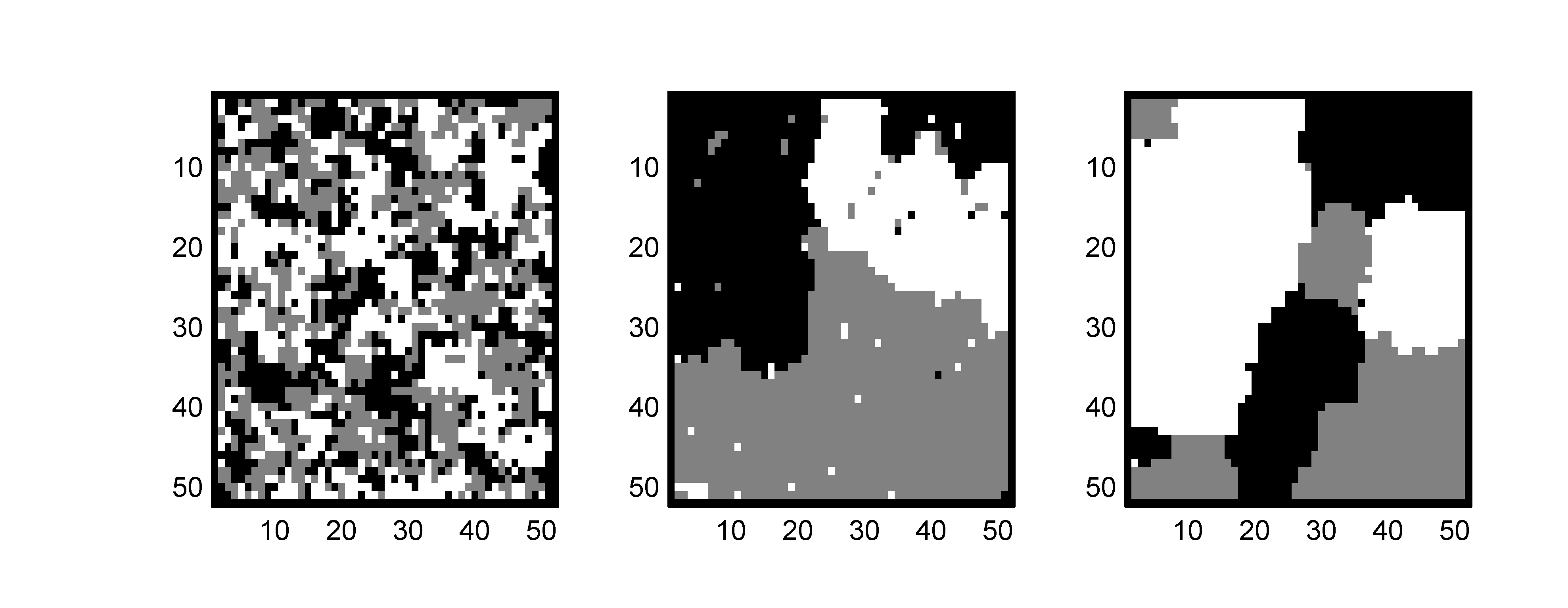}
\caption{Synthetic images generated from a Potts-Markov model with
(from left to right) $\GRAN = 0.8,\ 1.4,\  2$.}\label{fig:beta}
\end{figure}

Since the pioneer work of Geman \cite{Geman1984}, MRFs have been
widely used in the image processing community as in
\cite{Kevrann1995,Tonazzini2006}. The hyperspectral community has
also recently exploited the advantages of MRFs for hyperspectral
image analysis \cite{Rand2003,Rellier2004,Bali2008}. However, to our
knowledge, MRFs have not been studied for hyperspectral image
unmixing. MRFs provide an efficient way of modeling correlations
between pixels, which is adapted to the intrinsic properties of most
images. Two specific MRFs are appropriate for image analysis: the
Ising model for binary random variables and the Potts-Markov model
that is a simple generalization to more-than-two variables
\cite{Wu1982}. This paper focuses on the Potts-Markov model since it
is very appropriate to hyperspectral image segmentation
\cite{Bali2008}. Given a discrete random field $\boldsymbol{z}$
attached to an image with $P$ pixels, the Hammersley-Clifford
theorem yields
\begin{equation}
\label{eq:MRF}
 f\left(\LABELvec\right) =
\frac{1}{G(\GRAN)} \exp\left[\sum_{p=1}^P \sum_{p' \in
\mathcal{V}(p)} \GRAN \delta(\LABEL_p - \LABEL_{p'})\right]
\end{equation}
where $\GRAN$ is the \emph{granularity} coefficient, $G(\GRAN)$ is
the normalizing constant or \emph{partition function}
\cite{Kindermann1980} and $\delta(\cdot)$ is the Kronecker function
\begin{equation*} \delta(x)= \left\{
  \begin{array}{ll}
    1, & \textrm{if}~x = 0, \\
    0, &\textrm{otherwise}.
  \end{array}
\right.
\end{equation*}
Note that drawing a label vector
$\LABELvec=\left[\LABEL_1,\ldots,\LABEL_P\right]$ from the
distribution \eqref{eq:MRF} can be easily achieved without knowing
$G(\GRAN)$ by using a Gibbs sampler (the corresponding algorithmic
scheme is summarized in \cite{Eches2010_techreport_TGRS}). However,
a major difficulty with the distribution \eqref{eq:MRF} comes from
the partition function that has no closed-form expression and
depends on the unknown hyperparameter $\GRAN$. The hyperparameter
$\GRAN$ tunes the degree of homogeneity of each region in the image.
Some simulations have been conducted to show the influence of this
parameter on image homogeneity. Synthetic images have been generated
from a Potts-Markov model with $K=3$ (corresponding to three gray
levels in the image) and a $1$-order neighborhood structure. Fig.
\ref{fig:beta} indicates that a small value of $\GRAN$ induces a
\emph{noisy} image with a large number of regions, contrary to a
large value of $\GRAN$ that leads to few and large homogeneous
regions. It is unnecessary to consider values of $\GRAN \geq 2$
since for the $1$-order neighborhood
structure adopted here, ``\textit{When $\GRAN \geq 2$, the
Potts-Markov model is almost surely concentrated on single-color
images}'' \cite[p. 237]{Marin2007}. Note however that for larger
neighborhood systems, a smaller value of $\GRAN$ would be enough to
obtain uniform patches in Potts realizations since, for example,
$\GRAN$ is expected to be about twice for an $2$-order neighborhood
structure \cite{Ripley1988}. In this work, the granularity
coefficient $\GRAN$ will be fixed \emph{a priori}. However, it is
interesting to mention that the estimation of $\GRAN$ might also be
conducted by using the methods studied in \cite{Zhou1997},
\cite{Descombes1999} and \cite{Celeux2003}.


\subsection{Abundance Reparametrization}
\label{subsec:reparametrization} As explained before, the fraction
vectors $\ABUNDVEC_p$ should satisfy positivity and sum-to-one
constraints defined as
\begin{equation}\label{eq:constraints}
\left \{\begin{array}{l}
 \ABUND_r > 0,\forall r = 1,\ldots,R,\\
  \sum_{r=1}^R \ABUND_r = 1.
  \end{array}
  \right.
\end{equation}
To ensure that these abundance constraints are satisfied, we have
considered a reparametrization for positive parameters summing to
one that was introduced in \cite{Kent1988} for the spectral unmixing
of satellite images. Note that this reparametrization has also shown
interesting results for a pharmacokinetic problem \cite{Gelman1996}
and has been recently applied to hyperspectral unmixing
\cite{Themelis2008}. This reparametrization consists of rewriting
the abundances as a function of random variables that will be
referred to as \emph{logistic coefficients} in the rest of the
paper. A logistic coefficient vector $\COEFFVEC_p =
\left[\COEFF_{1,p}\ldots,\COEFF_{R,p} \right]^T$ is assigned to each
abundance vector $\ABUNDVEC_p$, according to the relationship
\begin{equation}\label{eq:abundance_reparam}
\ABUND_{r,p} = \frac{\exp(\COEFF_{r,p})}{\sum_{r=1}^R
\exp(\COEFF_{r,p})}.
\end{equation}
Initially, the spatial dependencies resulting from the image
partitioning described in Section \ref{subsec:spatial_dependencies}
are based on the first and second order moments of the abundance
vectors $\ABUNDVEC_p$. However, the spatial constraints defined in
\eqref{eq:class_statistics} can be easily adapted when using
logistic coefficient vectors. Indeed, in each class, the unknown
logistic coefficient vectors are assumed to share the same first and
second order moments, i.e., $\forall k\in\left\{1,\ldots,K\right\},\
\forall (p,p') \in \Ik\times \Ik$
\begin{equation}
\label{eq:class_statistics_bis}
\begin{split}
   \MCOEFFvec_k &= \mathrm{E}\left[\COEFFVEC_p\big|\LABEL_p=k\right] = \mathrm{E}\left[\COEFFVEC_{p'}\big|\LABEL_{p'}=k\right] \\
   \VCOEFFvec_k & =\mathrm{E}\left[\left(\COEFFVEC_p-\MCOEFFvec_k\right)\left(\COEFFVEC_p-\MCOEFFvec_k\right)\transp\big|\LABEL_p=k\right]\\
               &=
               \mathrm{E}\left[\left(\COEFFVEC_{p'}-\MCOEFFvec_k\right)\left(\COEFFVEC_{p'}-\MCOEFFvec_k\right)\transp\big|\LABEL_{p'}=k\right].
\end{split}
\end{equation}
With this reparametrization, the $k$th class is fully characterized
by the unknown hyperparameters $\MCOEFFvec_k$ and $\VCOEFFvec_k$.

\section{Hierarchical Bayesian model}
\label{sec:Bayesian_model} This section investigates the likelihood
and the priors inherent to the LMM for the spectral unmixing of
hyperspectral images, based on Potts-Markov  random fields and
logistic coefficients.

\subsection{Unknown parameters}
The unknown parameter vector associated to to the LMM unmixing
strategy is denoted as
\begin{equation*}
\BTHETA= \{\COEFFMAT, \LABELvec, \vnoise^2 \}
\end{equation*}
where $\vnoise^2$ is the noise variance, $\LABELvec$ is the label
vector and $\COEFFMAT = \left[\COEFFVEC_1,\ldots,\COEFFVEC_P
\right]$ with $\COEFFVEC_p = \left[\COEFFVEC_{1,p}, \ldots,
\COEFFVEC_{R,p} \right]^T~ (p = 1,\ldots,P)$ is the logistic
coefficient matrix used for the abundance reparametrization.
Note that the noise variance $\vnoise^2$ has been assumed to be unknown in the present paper,
contrary to the model considered in \cite{Kent1988}.

\subsection{Likelihood}
The additive white
Gaussian noise sequence of the LMM allows one to write\footnote{Note
that the dependence of the abundance vector $\ABUNDVEC_p$ on the
logistic coefficient vector $\COEFFVEC_p$ through
\eqref{eq:abundance_reparam} has been explicitly mentioned by
denoting $\ABUNDVEC_p=\ABUNDVEC_p(\COEFFVEC_p)$.}
$\Y_p|\COEFFVEC_p,\vnoise^2 \thicksim \mathcal{N}\left(\SIGNMAT
\ABUNDVEC_p(\COEFFVEC_p), \vnoise^2 \boldsymbol{I}_L\right)$ ($p =
1,\ldots,P$).  Therefore the likelihood function of $\Y_p$ is
\begin{equation}
f\left(\Y_p\,|\COEFFVEC_p,\vnoise^2\right) \propto
 \frac{1}{\vnoise^L} \exp \left[
 - \frac{\| \Y_p - \SIGNMAT \ABUNDVEC_p(\COEFFVEC_p) \|^2}{2
\vnoise^2}\right]
\end{equation}
where $\propto$ means proportional to and $\|\boldsymbol{x}\| =
\sqrt{\boldsymbol{x}^T \boldsymbol{x}}$ is the standard $\ell^2$
norm. By assuming independence between the noise sequences
$\boldsymbol{n}_p$ ($p = 1,\ldots,P$), the likelihood of the $P$
image pixels is
\begin{equation}
f\left(\YMAT|\COEFFMAT,\vnoise^2\right) = \prod_{p=1}^P
f\left(\Y_p|\COEFFVEC_p,\vnoise^2\right).
\end{equation}

\subsection{Parameter priors}
\begin{figure}[htbp]
\centering
\includegraphics[width=\figwidthsmall]{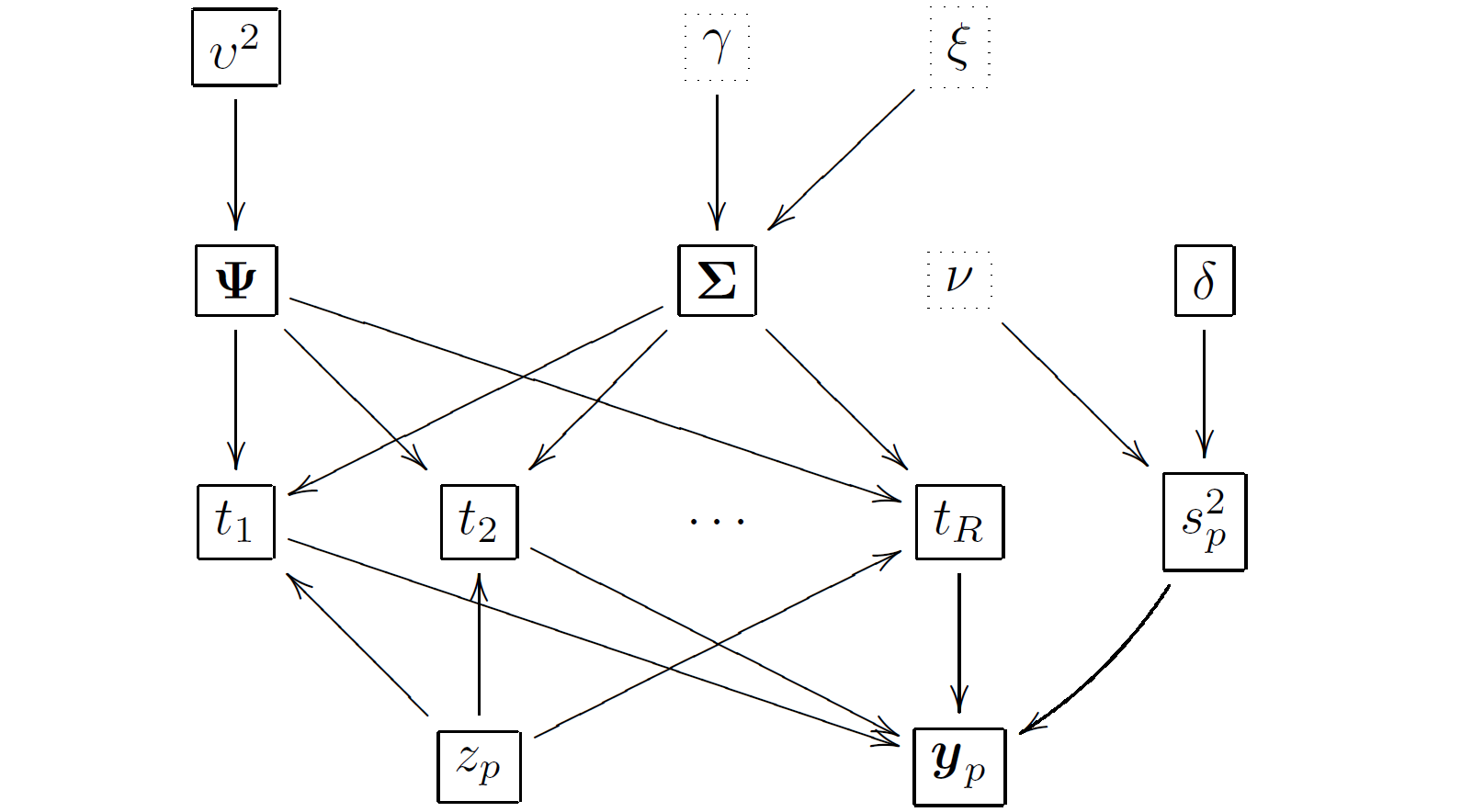}
\caption{DAG for the parameter priors and hyperpriors (the fixed
parameters appear in dashed boxes) for the LMM.}\label{fig:DAG}
\end{figure}

This section defines the prior distributions of the unknown
parameters and their associated hyperparameters that will be used
for the LMM. The directed acyclic graph (DAG) for the parameter
priors and hyperpriors for the considered model is represented in
Fig. \ref{fig:DAG}.

\subsubsection{Label prior}
\label{subsubsec:prior_label} The prior distribution for the label
vector $\LABELvec = \left[\LABEL_1,\ldots,\LABEL_P \right]^T$
introduced in paragraph \ref{subsec:MRF} is a Potts-Markov random
field with a $1$-order neighborhood and a known granularity
coefficient $\GRAN$ (fixed \emph{a priori}). The resulting prior
distribution can be written as in \eqref{eq:MRF} where
$\mathcal{V}(p)$ is the $1$-order neighborhood depicted in
Fig.~\ref{fig:clique} (left).

\subsubsection{Logistic coefficient prior}
Following the approach described in Section
\ref{subsec:spatial_dependencies}, each component of $\COEFFVEC_p$
is assumed to be distributed according to a Gaussian distribution.
In addition, as highlighted in \ref{subsec:reparametrization} (see
\eqref{eq:class_statistics_bis}), the mean  and variance of the
logistic coefficients depend on the class to which the corresponding
pixel belong. Therefore, the prior distribution for the
$\COEFFVEC_p$ is explicitly defined conditionally upon the pixel
label
\begin{equation}
\label{eq:prior_coeff}
 \COEFF_{r,p}|\LABEL_p = k,
\MCOEFF_{r,k}, \VCOEFF_{r,k} \thicksim
\mathcal{N}\left(\MCOEFF_{r,k},\VCOEFF_{r,k} \right)
\end{equation}
where the hyperparameters $\MCOEFF_{r,k}$ and $\VCOEFF_{r,k}$ depend
on the associated pixel class $k$. As suggested in Section
\ref{sec:introduction}, a hierarchical Bayesian algorithm will be
used to estimate these hyperparameters. For a given pixel $p$, by
assuming prior independence between the coefficients
$\COEFF_{1,p},\ldots,\COEFF_{R,p}$, the prior distribution for the
vector
$\COEFFVEC=\left[\COEFF_{1,p},\ldots,\COEFF_{R,p}\right]\transp$ is
\begin{equation}
f\left(\COEFFVEC_p|\LABEL_p = k, \MCOEFFvec_k,
\VCOEFFvec_k\right)\thicksim
\mathcal{N}\left(\MCOEFFvec_k,\VCOEFFvec_k \right)
\end{equation}
where
$\MCOEFFvec_k=\left[\MCOEFF_{1,k},\ldots,\MCOEFF_{R,k}\right]\transp$
and $\VCOEFFvec_k=\textrm{diag}\left(\VCOEFF_{r,k}\right)$ is the
$R\times R$ diagonal matrix whose diagonal elements are
$\VCOEFF_{r,k}$.

By assuming prior independence between the $P$ vectors
$\COEFFVEC_1,\ldots,\COEFFVEC_P$, the full posterior distribution
for the logistic coefficient matrix $\COEFFMAT$ is
\begin{equation}\label{eq:prior_coeff_full}
f\left(\COEFFMAT | \LABELvec, \GMCOEFF, \GVCOEFF\right) =
\prod_{k=1}^K \prod_{p \in \Ik} f\left(\COEFFVEC_p|\LABEL_p = k,
\MCOEFFvec_k, \VCOEFFvec_k\right)
\end{equation}
with $\GMCOEFF = \left[\MCOEFFvec_1,\ldots,\MCOEFFvec_K \right]$ and
$\GVCOEFF = \left\{\VCOEFFvec_1,\ldots,\VCOEFFvec_K \right\}$.

\subsubsection{Noise variance prior}
 A conjugate inverse-gamma
distribution is assigned to the noise variance
\begin{equation}
    \vnoise^2 | \nu, \HyperVAR \thicksim \mathcal{IG}(\nu,\HyperVAR)
\end{equation}
where $\nu$ and $\HyperVAR$ are adjustable hyperparameters. This
paper assumes $\nu=1$ (as in \cite{Punskaya2002} or
\cite{Tourneret2007}) and estimates $\HyperVAR$ jointly with the
other unknown parameters and hyperparameters (using a hierarchical
Bayesian algorithm).

\subsection{Hyperparameter priors}
Hierarchical Bayesian algorithms require to define prior
distributions for the hyperparameters. A particular attention has to
be devoted to the hyperparameters $\MCOEFF_{r,k}$ and
$\VCOEFF_{r,k}$ since they fully describe the different classes
partitioning the image. The prior distributions for $\MCOEFF_{r,k}$
and $\VCOEFF_{r,k}$ are conjugate distributions. More precisely, a
vague inverse-gamma distribution is chosen for the logistic
coefficient variance $\VCOEFF_{r,k}$, i.e.,
\begin{equation}
\VCOEFF_{r,k} | \HyperVCOEFFa, \HyperVCOEFFb \thicksim
\mathcal{IG}(\HyperVCOEFFa,\HyperVCOEFFb)
\end{equation}
where $\HyperVCOEFFa$ and $\HyperVCOEFFb$ have been tuned to
$\HyperVCOEFFa=1$ and $\HyperVCOEFFb=5$ (in order to obtain a large
variance). Moreover, a centered Gaussian distribution with unknown
variance has been chosen as prior for the logistic coefficient mean
\begin{equation}
\MCOEFF_{r,k} | \HyperMCOEFF \thicksim
\mathcal{N}\left(0,\HyperMCOEFF \right)
\end{equation}
where $\HyperMCOEFF$ is another adjustable hyperparameter. By
assuming independence between the different mean vectors
$\MCOEFFvec_k$, as well as between the covariance matrices
$\VCOEFFvec_k$ for $k = 1, \ldots, K$, the full priors for the two
hyperparameters $\GMCOEFF$ and $\GVCOEFF$ can be expressed as
\begin{equation}\label{eq:prior_mcoeff}
f(\GMCOEFF | \HyperMCOEFF) \propto \prod_{k=1}^K \prod_{r=1}^R
\left(\frac{1}{\HyperMCOEFF} \right)^{\frac{1}{2}}
\exp\left(-\frac{\MCOEFF_{r,k}^2}{2\HyperMCOEFF}\right)
\end{equation}
\begin{equation}\label{eq:prior_vcoeff}
f(\GVCOEFF | \HyperVCOEFFa, \HyperVCOEFFb) \propto \prod_{k=1}^K
\prod_{r=1}^R
\frac{\HyperVCOEFFb^\HyperVCOEFFa}{\Gamma(\HyperVCOEFFa)}
(\VCOEFF_{r,k})^{-(\HyperVCOEFFa+1)}\exp\left(-\frac{\HyperVCOEFFb}{\VCOEFF_{r,k}}\right).
\end{equation}
Jeffreys' priors are chosen for the hyperparameters $\HyperVAR$ and
$\HyperMCOEFF$ (see, e.g., \cite[p.
131]{Robert2007choice} for details including computations)
\begin{equation}\label{eq:hyper_prior}
f(\HyperVAR) \propto
\frac{1}{\HyperVAR}\mathbf{1}_{\mathbb{R}^+}(\HyperVAR), \quad
f(\HyperMCOEFF) \propto
\frac{1}{\HyperMCOEFF}\mathbf{1}_{\mathbb{R}^+}(\HyperMCOEFF).
\end{equation}
where $\mathbf{1}_{\mathbb{R}^+}(\cdot)$ denotes the indicator
function defined on $\mathbb{R}^+$. These choices, also adopted in
\cite{Punskaya2002,Dobigeon_IEEE_TSP_2007b}, reflect the lack of
knowledge regarding these two hyperparameters. At this last
hierarchy level within the Bayesian inference, the hyperparameter
vector can be defined as $\HyperMAT =
\left\{\GMCOEFF,\GVCOEFF,\HyperMCOEFF,\HyperVAR \right\}$.

\subsection{Joint distribution}
The joint posterior distribution of the unknown parameters and
hyperparameters is classically defined using the hierarchical
structure
\begin{equation}
f(\BTHETA,\HyperMAT|\YMAT)  =
f(\YMAT|\BTHETA)f(\BTHETA|\HyperMAT)f(\HyperMAT).\\
\end{equation}
Straightforward computations yield the following posterior
\begin{equation}\label{eq:posterior_LMM}
\begin{split}
&f(\BTHETA,\HyperMAT| \YMAT) \propto
\left(\frac{1}{\vnoise^2} \right)^{\frac{L P}{2}} \prod_{p=1}^P \exp \left[ -
\frac{\| \Y_p - \SIGNMAT \ABUNDVEC_p(\COEFFVEC_p) \|^2}{2
\vnoise^2}\right] \\
&\times \exp\left[\sum_{p=1}^P \sum_{p' \in \mathcal{V}(p)} \GRAN \delta(\LABEL_p - \LABEL_{p'})\right]\\
&\times \frac{\HyperVAR^{\nu-1}}{\left(\vnoise^{2}\right)^{\nu+1}}\exp\left({-\frac{\HyperVAR}{\vnoise^2}}\right) \prod_{p=1}^P \left(\frac{1}{\HyperMCOEFF}\right)^{\frac{RK}{2}+1} \\
& \times   \prod_{r,k}   \frac{1}{\sigma_{r,k}^{n_k+1}} \exp\left[-
\left( \frac{\MCOEFF_{r,k}^2}{2\HyperMCOEFF} +
\frac{2\HyperVCOEFFb+\sum_{p \in
    \mathcal{I}_k}(\COEFF_{r,p} -
    \MCOEFF_{r,k})^2}{2\VCOEFF_{r,k}}\right) \right]
\end{split}
\end{equation}
with $n_k = \textrm{card}(\Ik)$. The posterior distribution
\eqref{eq:posterior_LMM} associated to the LMM is too complex to
obtain closed-from expressions for the MMSE or MAP estimators of the
unknown parameter vector $\BTHETA$. To alleviate this problem, we
propose to use MCMC methods to generate samples that are
asymptotically distributed according to \eqref{eq:posterior_LMM}.
The generated samples are then used to approximate the Bayesian
estimators. The next section studies a hybrid Gibbs sampler that
generates samples asymptotically distributed according to the
posterior distribution \eqref{eq:posterior_LMM}.

\section{Hybrid Gibbs sampler}\label{sec:hybrid_gibbs_sampler}
This section studies a Metropolis-within-Gibbs sampler that
generates samples according to the joint posterior
$f(\BTHETA,\HyperMAT| \YMAT)$. The proposed sampler iteratively
generates samples distributed according to the conditional
distributions detailed below.

\subsection{Conditional distribution of the label vector $\LABELvec$}
For each pixel $p~(p = 1,\ldots,P)$, the class label $\LABEL_p$ is a
discrete random variable whose conditional distribution is fully
characterized by the probabilities
\begin{equation}
\label{eq:label_post_ini} \mathrm{P}\left[\LABEL_p = k
|\LABELvec_{\textrm{-}p},\COEFFVEC_p,\MCOEFFvec_k,\VCOEFFvec_k
\right] \propto f(\COEFFVEC_p|\LABEL_p = k,
\MCOEFFvec_k,\VCOEFFvec_k)f\left(\LABEL_p|\LABELvec_{\textrm{-}p}\right)
\end{equation}
where $k=1,...,K$ ($K$ is the number of classes) and
$\LABELvec_{\textrm{-}p}$ denotes the vector $\LABELvec$ whose $p$th
element has been removed.  These posterior probabilities can be
expressed as
\begin{equation}\label{eq:label_post}
\begin{split}
\mathrm{P}&\left[\LABEL_p = k
|\LABELvec_{\textrm{-}p},\COEFFVEC_p,\MCOEFFvec_k,\VCOEFFvec_k
\right] \\
&\propto  \exp\left[\sum_{p=1}^P \sum_{p' \in \mathcal{V}(p)} \GRAN
\delta(\LABEL_p - \LABEL_{p'})\right] \\
&\times \left|\VCOEFFvec_k\right|^{-1/2} \exp\left[-\frac{1}{2}
\left(\COEFFVEC_p - \MCOEFFvec_k\right)^T
\VCOEFFvec_k^{-1}\left(\COEFFVEC_p - \MCOEFFvec_k\right) \right]
\end{split}
\end{equation}
where $\left|\VCOEFFvec_k\right| = \prod_{r=1}^R \VCOEFF_{r,k}$.
Note that the posterior probabilities of the label vector
$\LABELvec$ in \eqref{eq:label_post} define an MRF. Consequently,
sampling from this conditional distribution can be achieved using
the scheme detailed in \cite{Eches2010_techreport_TGRS}, i.e., by
drawing a discrete value in the finite set
$\left\{1,\ldots,K\right\}$ with the probabilities
\eqref{eq:label_post}.

\subsection{Conditional distribution of logistic coefficient matrix $\COEFFMAT$}
For each pixel $p$, the Bayes theorem yields
\begin{equation*}
f\left(\COEFFVEC_p | \LABEL_p = k, \MCOEFFvec_k,\VCOEFFvec_k,\Y_p
\right) \propto f\left(\Y_p | \COEFFVEC_p, \vnoise^2\right)f\left(\COEFFVEC_p
|\LABEL_p = k, \MCOEFFvec_k,\VCOEFFvec_k \right).
\end{equation*}
\noindent Straightforward computations lead to
\begin{equation}\label{eq:posterior_coef}
\begin{split}
f&\left(\COEFFVEC_p | \LABEL_p = k, \MCOEFFvec_k,
    \VCOEFFvec_k,\Y_p,\vnoise^2 \right)\\
&\propto     \left(\frac{1}{\vnoise^2}\right)^{\frac{L}{2}}
\exp\left\{-\frac{1}{2\vnoise^2}\norm{\Y_p -
    \SIGNMAT \ABUNDVEC_p(\COEFFVEC_p)}^2  \right\}\\
&\times \left|\VCOEFFvec_k\right|^{-\frac{1}{2}} \exp
\left[-\frac{1}{2}
    \left(\COEFFVEC_p - \MCOEFFvec_k\right)^T
    \VCOEFFvec_k^{-1}\left(\COEFFVEC_p - \MCOEFFvec_k\right) \right].
\end{split}
\end{equation}
Unfortunately, it is too difficult to generate samples distributed
according to \eqref{eq:posterior_coef}. Therefore, a
Metropolis-Hastings step is used, based on a random walk method
\cite[p. 245]{Robert2004} with a Gaussian distribution
$\mathcal{N}(0,\vrnwalk_r^2)$ as proposal distribution. The variance
$\vrnwalk_r^2$ of the instrumental distribution has been fixed to
obtain an acceptance rate between $0.15$ and $0.5$  as recommended
in \cite{Roberts1996}.

\subsection{Conditional distributions of the noise variance}
The
Bayes theorem yields
\begin{equation*}
f\left(\vnoise^2|\YMAT,\COEFFMAT,\HyperVAR\right) \propto f\left(\vnoise^2|\HyperVAR\right) \prod_{p=1}^P f(\Y_p
\,|\COEFFVEC_p,\vnoise^2) .
\end{equation*}
As a consequence, $\vnoise^2|\YMAT,\COEFFMAT,\HyperVAR$ is
distributed according to the inverse-Gamma distribution
\begin{equation}\label{eq:posterior_noise}
\vnoise^2 |\YMAT,\COEFFMAT, \HyperVAR \thicksim
\mathcal{IG}\left(\frac{L P}{2} + 1, \HyperVAR + \sum_{p=1}^P \frac{\| \Y_p - \SIGNMAT
\ABUNDVEC_p(\COEFFVEC_p) \|^2}{2} \right).
\end{equation}

\subsection{Conditional distribution of $\GMCOEFF$ and $\GVCOEFF$}
For each endmember $r$ ($r = 1,\ldots, R$) and each class $k$
($k=1,\ldots,K$), the conditional distribution of $\MCOEFF_{r,k}$
can be written as
\begin{multline}
f\left(\MCOEFF_{r,k} | \LABELvec, \COEFFVEC_r, \VCOEFF_{r,k},
\HyperMCOEFF \right)\\ \propto  f\left(\MCOEFF_{r,k} |\HyperMCOEFF
\right) \prod_{p \in \Ik} f\left(\COEFF_{r,p}|z_p = k,
\MCOEFF_{r,k}, \VCOEFF_{r,k}\right).
\end{multline}
Similarly, the conditional distribution of $\VCOEFF_{r,k}$ is
\begin{equation} f\left(\VCOEFF_{r,k} |
\LABELvec, \COEFFVEC_r,  \MCOEFF_{r,k} \right) \propto
f\left(\VCOEFF_{r,k}\right) \prod_{p \in \mathcal{I}_k}
f\left(\COEFF_{r,p}|\LABEL_p = k, \MCOEFF_{r,k},
\VCOEFF_{r,k}\right).
\end{equation}
Straightforward computations allow one to obtain the following
results
\begin{equation}\label{eq:posterior_mcoeff}
\MCOEFF_{r,k} | \LABELvec, \COEFFVEC_r, \VCOEFF_{r,k}, \HyperMCOEFF
\thicksim \mathcal{N}\left(\frac{\HyperMCOEFF n_k
\overline{\COEFF}_{r,k}}{\VCOEFF_{r,k} + \HyperMCOEFF n_k},
\frac{\HyperMCOEFF \VCOEFF_{r,k}}{\VCOEFF_{r,k} + \HyperMCOEFF n_k}
\right)
\end{equation}
\begin{equation}\label{eq:posterior_vcoeff}
\VCOEFF_{r,k} | \LABELvec, \COEFFVEC_r, \MCOEFF_{r,k} \thicksim
\mathcal{IG}\left(\frac{n_k}{2} + 1,\HyperVCOEFFb + \sum_{p \in
\Ik}\frac{\left(\COEFF_{r,p} - \MCOEFF_{r,k} \right)^2}{2}\right)
\end{equation}
with $\overline{\COEFF}_{r,k} = \frac{1}{n_k} \sum_{p \in \Ik}
\COEFF_{r,p}$

\subsection{Conditional distribution of $\HyperMCOEFF$ and $\HyperVAR$}
The conditional distributions of $\HyperMCOEFF$ and $\HyperVAR$ are
the following inverse-gamma and gamma distributions, respectively
\begin{equation*}
\HyperMCOEFF | \GMCOEFF \thicksim \mathcal{IG}\left(\frac{RK}{2},
\frac{1}{2}\sum_{k=1}^K \MCOEFFvec_{k}^T \MCOEFFvec_{k} \right),
\quad \HyperVAR | \vnoise^2 \thicksim \mathcal{G}\left(1,
\frac{1}{\vnoise^2} \right).
\end{equation*}

\section{Simulation results on synthetic data}\label{sec:simus_synthetic}


\begin{table}[htb]\small
\begin{center}
\renewcommand{\arraystretch}{1.3}
\caption{Actual and estimated abundance mean and variance in each
class.}
\begin{tabular}{|c|c|c|c|}
\cline{3-4}  \multicolumn{2}{c|}{} & Actual values & Estimated values \\
\hline \multirow{2}*{Class $1$} & $\boldsymbol{\mu}_1 =
\textrm{E}[\ABUNDVEC_{p,~p \in
\mathcal{I}_1}]$ & $[0.6, 0.3, 0.1]^T$ & $[0.57, 0.3, 0.13]^T$ \\
\cline{2-4} & $\textrm{Var}[\ABUND_{p,r,~p \in
\mathcal{I}_1}]~(\times 10^{-3})$ & $[5, 5, 5]^T$ & $[5.6, 6.7,
6.7]^T$ \\
\hline \hline \multirow{2}*{Class $2$} & $\boldsymbol{\mu}_2 =
\textrm{E}[\ABUNDVEC_{p,~p \in
\mathcal{I}_2}]$ & $[0.3, 0.5, 0.2]^T$ & $[0.29, 0.49, 0.2]^T$ \\
\cline{2-4} & $\textrm{Var}[\ABUND_{p,r,~p \in
\mathcal{I}_2}]~(\times 10^{-3})$ & $[5, 5, 5]^T$ & $[4.5, 5.2,
8.1]^T$\\
\hline \hline \multirow{2}*{Class $3$} & $\boldsymbol{\mu}_3 =
\textrm{E}[\ABUNDVEC_{p,~p \in
\mathcal{I}_3}]$ & $[0.3, 0.2, 0.5]^T$ & $[0.3, 0.2, 0.5]^T$ \\
\cline{2-4} & $\textrm{Var}[\ABUND_{p,r,~p \in
\mathcal{I}_3}]~(\times 10^{-3})$ & $[5, 5, 5]^T$ & $[4.6, 5.7, 10.2]^T$\\
\hline
\end{tabular}
\label{tab:Actual}
\end{center}
\end{table}

\begin{figure}[htbp]
\centering
\includegraphics[width=\figwidthsmall]{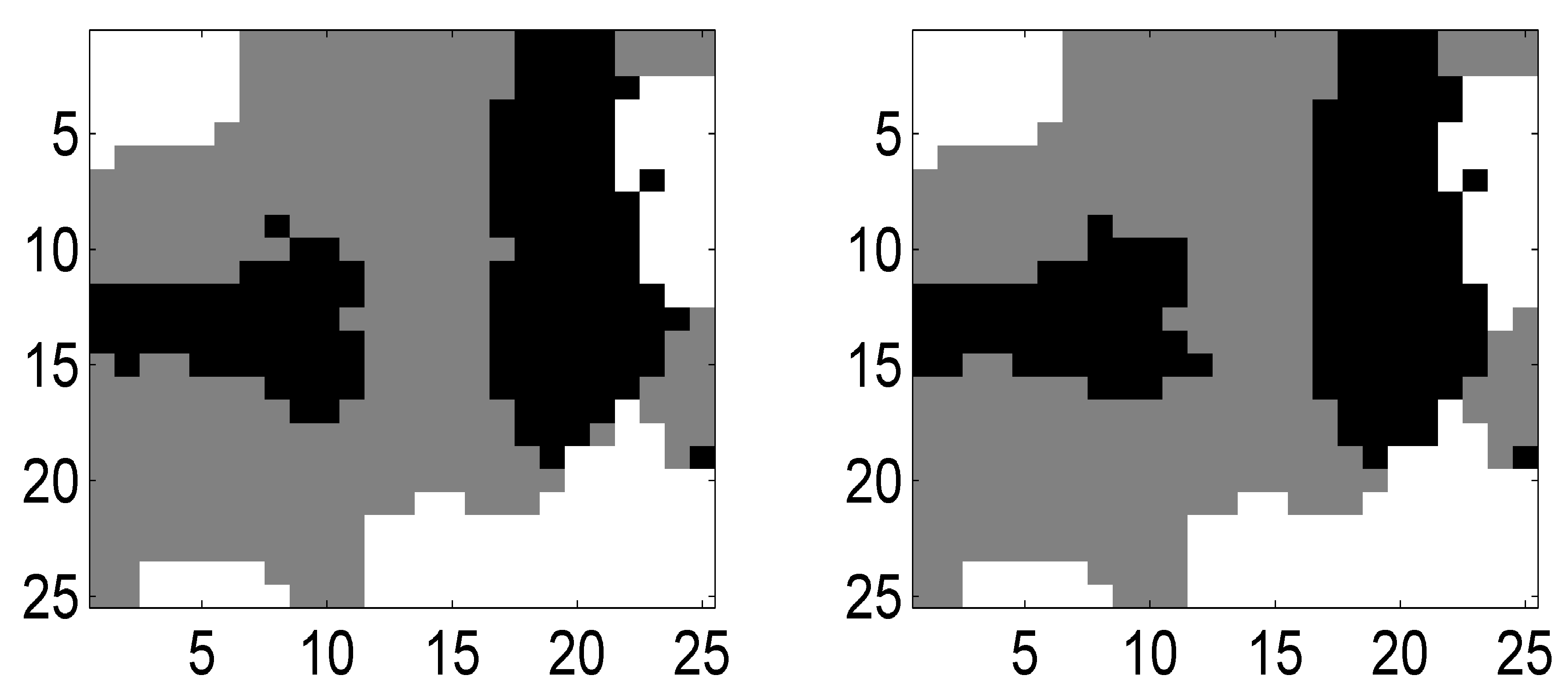}
\caption{Left: the actual label map. Right: the label map estimated
by the LMM hybrid Gibbs sampler. }\label{fig:labels_image}
\end{figure}

Many simulations have been conducted to illustrate the accuracy of
the proposed algorithm. The first experiment considers a $25 \times
25$ synthetic image with $K = 3$ different classes. The image
contains $R = 3$ mixed components (construction concrete, green
grass and micaceous loam) whose spectra ($L = 413$ spectral bands)
have been extracted from the spectral libraries distributed with the
ENVI package \cite{ENVImanual2003}. A label map shown in Fig.
\ref{fig:labels_image} (left) has been generated using
\eqref{eq:MRF} with $\GRAN = 1.1$.

\begin{figure}[htbp]
\centering
\includegraphics[width=\figwidth]{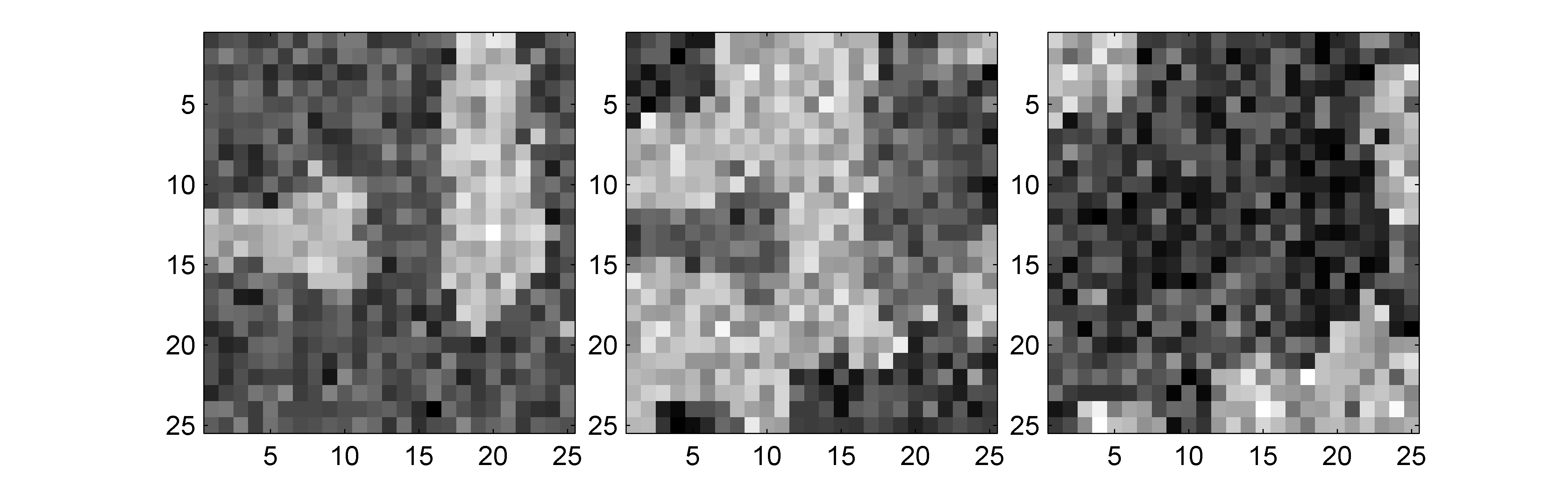}
\includegraphics[width=\figwidth]{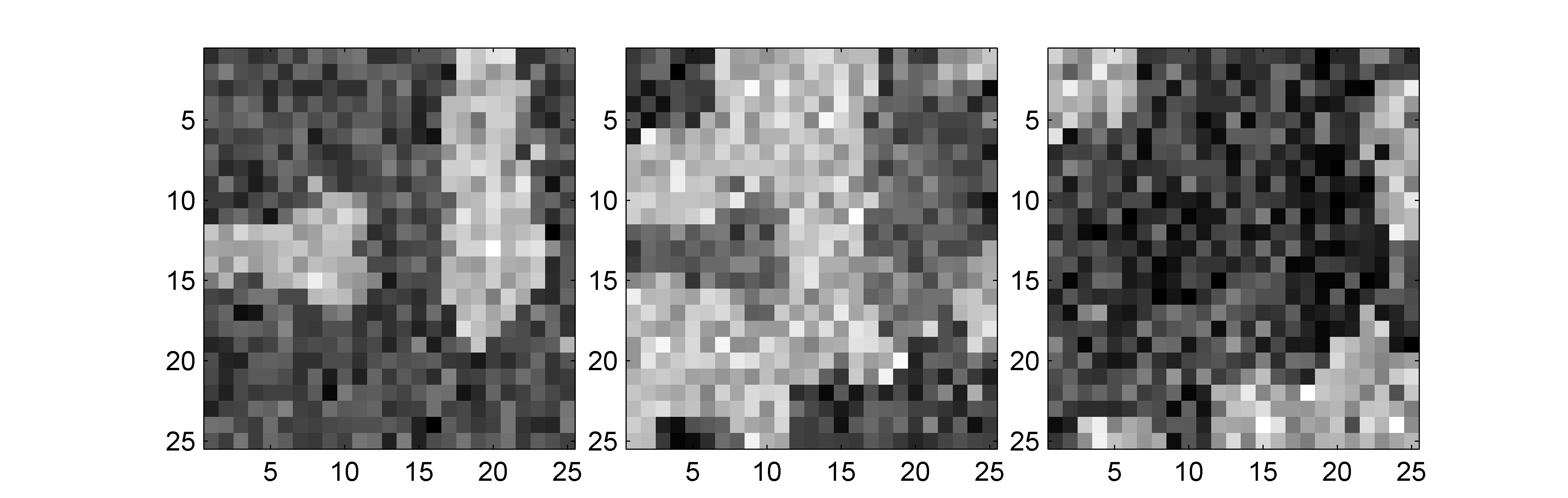} \caption{Top: abundance maps of the $3$ pure materials for LMM.
 Bottom: abundance maps of the $3$ pure materials estimated by the hybrid Gibbs sampler (from left to right: construction concrete,
green grass, micaceous loam).}\label{fig:synthetic_LMM_images}
\end{figure}

The mean and variance of the abundances have been chosen for each
class as reported in Table \ref{tab:Actual}. These values reflect
the fact that the $1$st endmember is more present in Class $1$ (with
average concentration of $60\%$), the $2$nd endmember is more
present in Class $2$ (with average concentration of $50\%$) and the
$3$rd endmember is more present in Class $3$ (with average
concentration of $50\%$). In this simulation scenario, the abundance
variance has been fixed to a common value $0.005$ for all
endmembers, pixels and classes. The generated abundance maps for the
LMM are depicted in Fig. \ref{fig:synthetic_LMM_images}. Note that a
white (resp. black) pixel in the fraction map indicates a large
(resp. small) value of the abundance coefficient.
The noise variance is chosen such as
the average signal-to-noise ratio (SNR) is equal to $\textrm{SNR} =
19 \textrm{dB}$, i.e. $\vnoise^2 = 0.001$.

\begin{figure}[htbp]
\centering
\includegraphics[width=\figwidth]{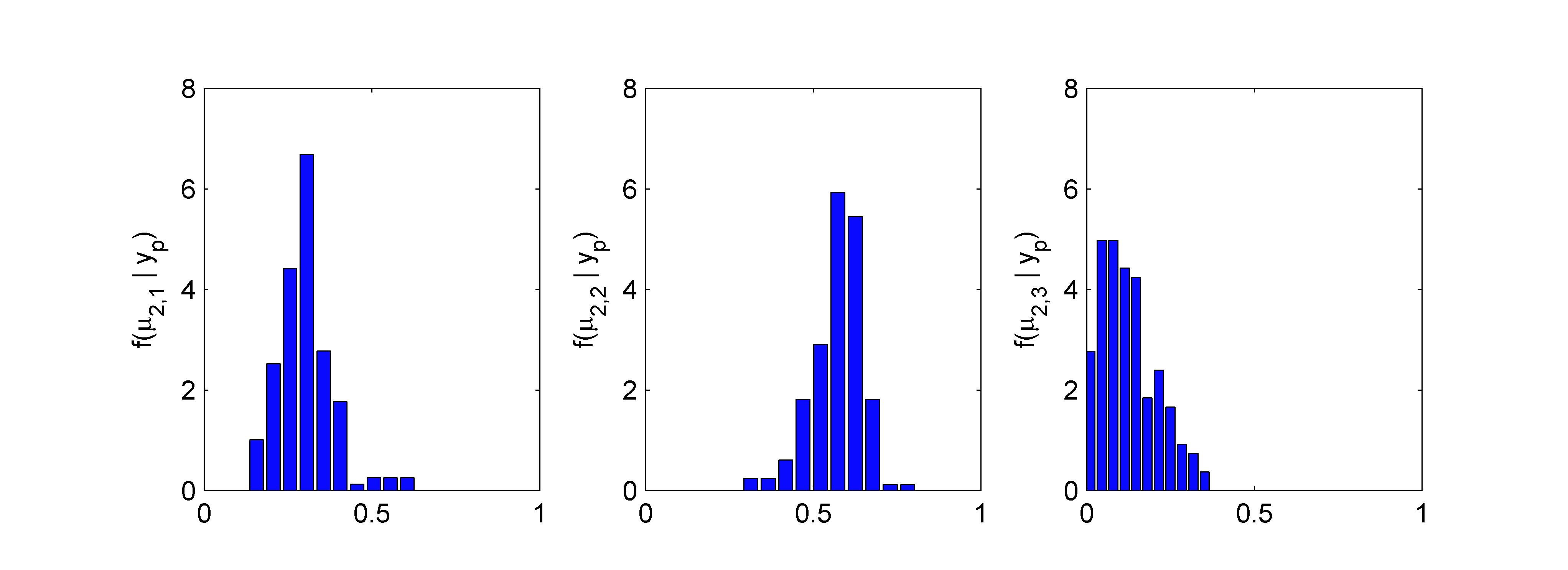}
\caption{Histograms of the abundance means $\boldsymbol{\mu}_k =
\left[\mu_{k,1},\mu_{k,2},\mu_{k,3}\right]\transp$ estimated by the
proposed hybrid Gibbs algorithm for the $2$nd class
($k=2$).}\label{fig:histo}
\end{figure}

The MMSE and MAP estimators for the unknown parameters can be
computed from samples generated with the Gibbs samplers presented in
Section~\ref{sec:hybrid_gibbs_sampler}. For instance, the marginal
MAP estimates of the label vector $\hat{\LABELvec}_{\textrm{MAP}}$
are depicted in Fig. \ref{fig:labels_image} (right) for the proposed
hybrid Gibbs algorithm. The MMSE estimates of the abundances
conditioned upon $\hat{\LABELvec}_{\textrm{MAP}}$ are shown in Fig.
\ref{fig:synthetic_LMM_images}. A number of $N_{\textrm{MC}} = 5000$
iterations (including $500$ burn-in iterations) has been necessary
to obtain these results. The proposed algorithm generates samples
distributed according to the full posterior of interest. Then, these
samples can be used to compute, for instance, the posterior
distributions of the mean vectors
$\boldsymbol{\mu}_k=\mathrm{E}\left[\ABUNDVEC_p\right]$
($k=1,\ldots,K$, $p\in\mathcal{I}_k$). These mean vectors,
introduced in \eqref{eq:class_statistics}, are of great interest
since they characterize each class. Therefore, as an additional
insight, the histograms of the abundance means $\boldsymbol{\mu}_k$
estimated by the proposed algorithm have been depicted in
Fig.~\ref{fig:histo} for the $2$nd class, i.e., $k=2$. Similar
results have been obtained for the other classes. They are omitted
here for brevity. Finally, the estimated abundance means and
variances have been reported in Table~\ref{tab:Actual} (last row).
The estimated classes, abundance coefficients and abundance mean
vectors are clearly in accordance with their actual values.

The LMM hybrid Gibbs algorithm is compared respectively with its
non-spatial constrained Bayesian counterpart developed in
\cite{Dobigeon_IEEE_TSP_2008}. The synthetic image shown in Fig.
\ref{fig:labels_image} has been analyzed by the initial algorithm of
\cite{Dobigeon_IEEE_TSP_2008} with the same number of iterations
$N_{\textrm{MC}}$ in addition with the FCLS \cite{Heinz2001}
algorithm. As a criterion, the global mean square error (MSE) of the
$r$th estimated abundances have been computed for each algorithm.
This global MSE is defined as
\begin{equation}
  \textrm{MSE}^2_r = \frac{1}{P}\sum_{p=1}^P (\hat{a}_{r,p} -
  a_{r,p})^2
\end{equation}
where $\hat{a}_{r,p}$ denotes the MMSE estimate of the abundance
$a_{r,p}$. Table \ref{tab:MSE} reports the different results showing
that the algorithm developed in this paper (referred to as
``Spatial") performs better than the non-spatial constrained
algorithms (referred to as ``Bayesian" and ``FCLS").

\begin{table}[htbp]
\renewcommand{\arraystretch}{1.4}
\begin{center}
\caption{Global MSEs of each abundance component.}
\begin{tabular}{|c|c|c|c|}
\cline{2-4} \multicolumn{1}{c|}{} & FCLS & Bayesian & Spatial \\
\hline $\textrm{MSE}^2_1$ & $0.0019$ & $0.0016$ & $3.1 \times 10^{-4}$ \\
\hline $\textrm{MSE}^2_2$ & $4.3 \times 10^{-4}$ & $4.1 \times 10^{-4}$ & $8.98 \times 10^{-5}$ \\
\hline $\textrm{MSE}^2_3$ & $0.0014$ & $0.0013$ & $2.35 \times 10^{-4}$ \\
\hline
\end{tabular}
\label{tab:MSE}
\end{center}
\end{table}

\section{Simulation results on  AVIRIS Images}\label{sec:simus_real}

\subsection{Performance of the proposed algorithm}

\begin{figure}[h!]
\centering
\includegraphics[width=\figwidthsmall]{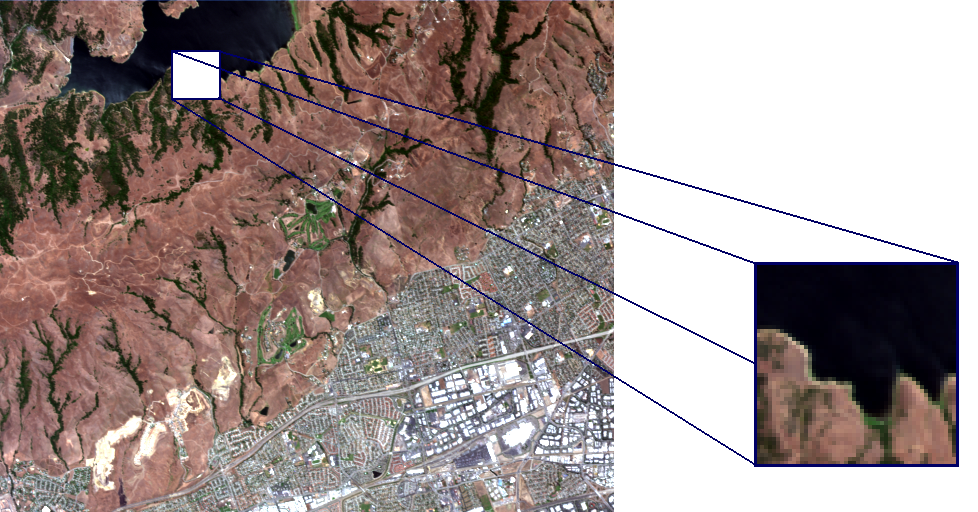}
\caption{Real hyperspectral data: Moffett field acquired by AVIRIS
in $1997$ (left) and the region of interest shown in true colors
(right).}\label{fig:Moffet ROI}
\end{figure}

\begin{figure}[htbp]
\centering
\includegraphics[width=\figwidth]{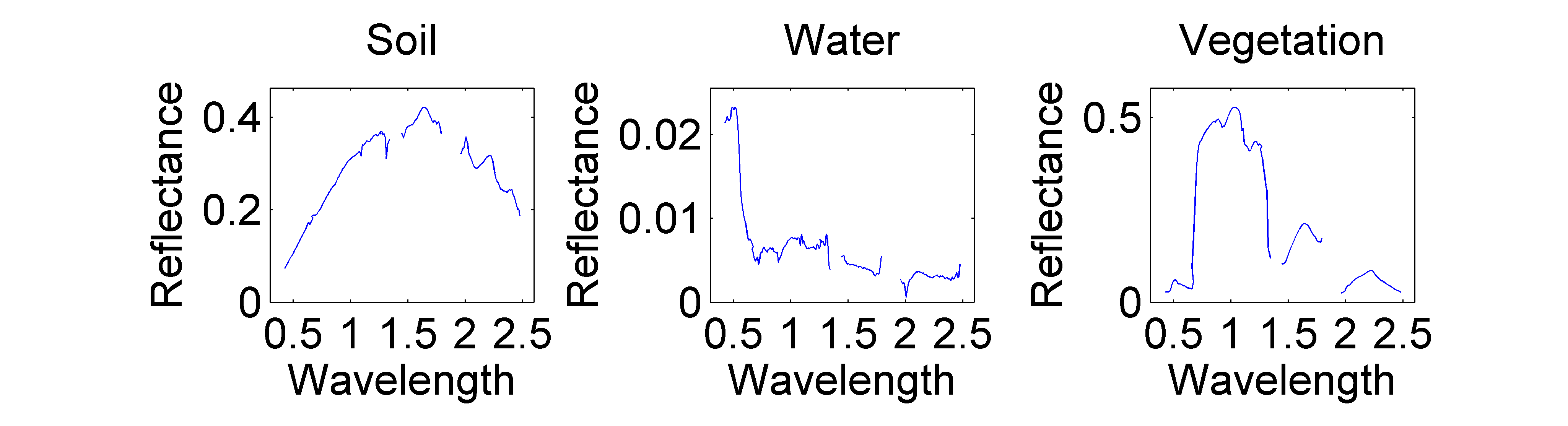}
\caption{The $R=3$ endmember spectra obtained by the N-FINDR
algorithm.}\label{fig:spectra_extracted_NFINDR}
\end{figure}

This section illustrates the performance of the proposed spatial
algorithm on a real hyperspectral dataset, acquired over Moffett
Field (CA, USA) in $1997$ by the JPL spectro-imager AVIRIS. Many
previous works have used this image to illustrate and compare
algorithm performance with hyperspectral images
\cite{Christophe2005,Akgun2005}. The
first region of interest, represented in Fig.~\ref{fig:Moffet ROI},
is a $50 \times 50$ image. The data set has been reduced from the
original $224$ bands to $L = 189$ bands by removing water absorption
bands. As in \cite{Dobigeon_IEEE_TSP_2008}, a principal component
analysis has been conducted as a processing step to determine the
number of endmembers present in the scene. Then, the endmembers
spectra have been extracted with the help of the endmember
extraction procedure N-FINDR proposed by Winter in
\cite{Winter1999}. The $R = 3$ extracted endmembers, shown in Fig.
\ref{fig:spectra_extracted_NFINDR}, corresponds to soil, vegetation
and water\footnote{Note that the influence of the endmember
extraction step on the unmixing results has been investigated in
\cite{Eches2010_techreport_TGRS} by coupling the proposed algorithm
with other EEAs.}. The algorithm proposed in Section
\ref{sec:hybrid_gibbs_sampler} has been applied on this image with
$N_{\textrm{MC}} = 5000$ iterations (with $500$ burn-in iterations).
The number of classes has been fixed to $K = 4$ since prior
knowledge on the scene allows one to identify $4$ areas in the
image: water point, lake shore, vegetation and soil.

\begin{figure}[h!]
\centering
\includegraphics[width=\figwidth]{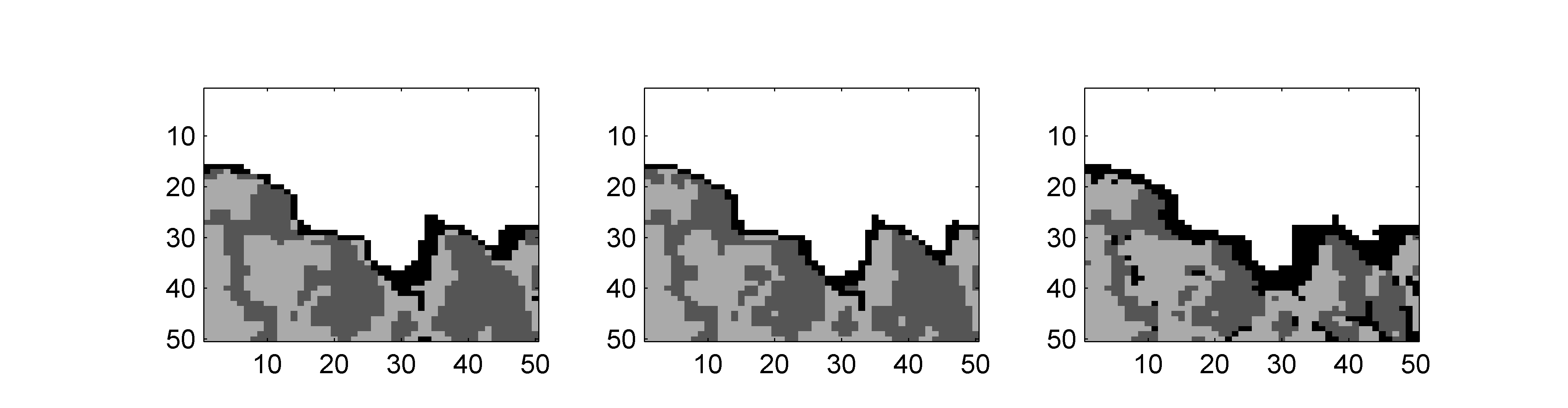}
\caption{Label map estimated by the LMM-based proposed algorithm for
$R = 3$ (left), $R = 4$ (middle) and $R = 5$
(right).}\label{fig:results_moffett_label}
\end{figure}

\begin{figure}[h!]
\centering
\includegraphics[width=\figwidth]{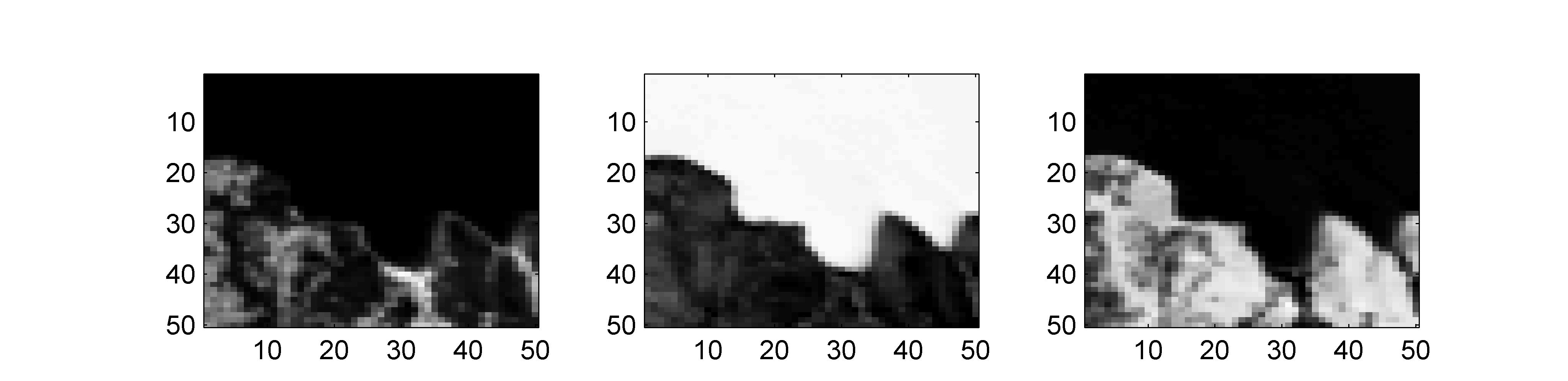}
\includegraphics[width=\figwidth]{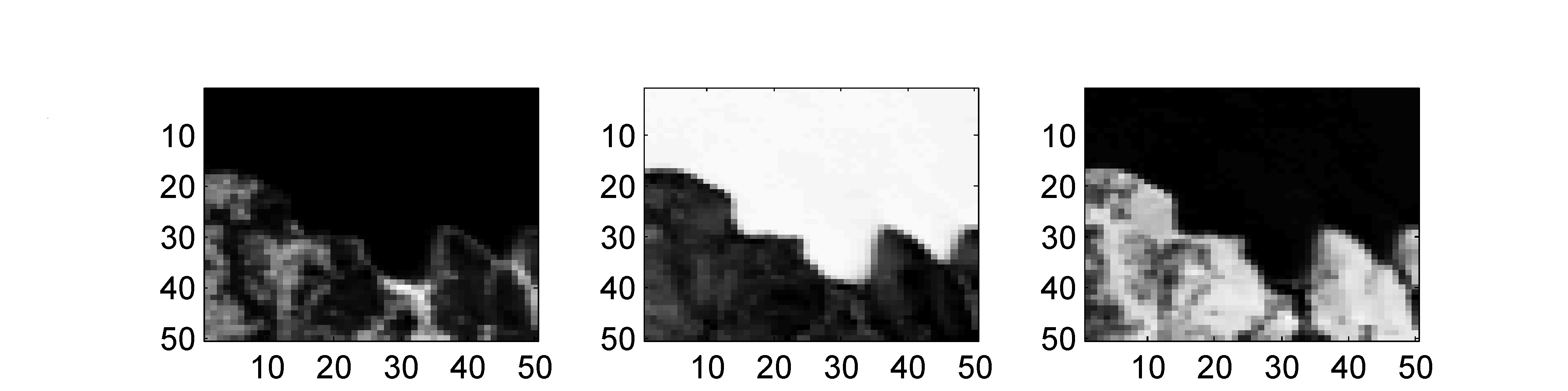}
\includegraphics[width=\figwidth]{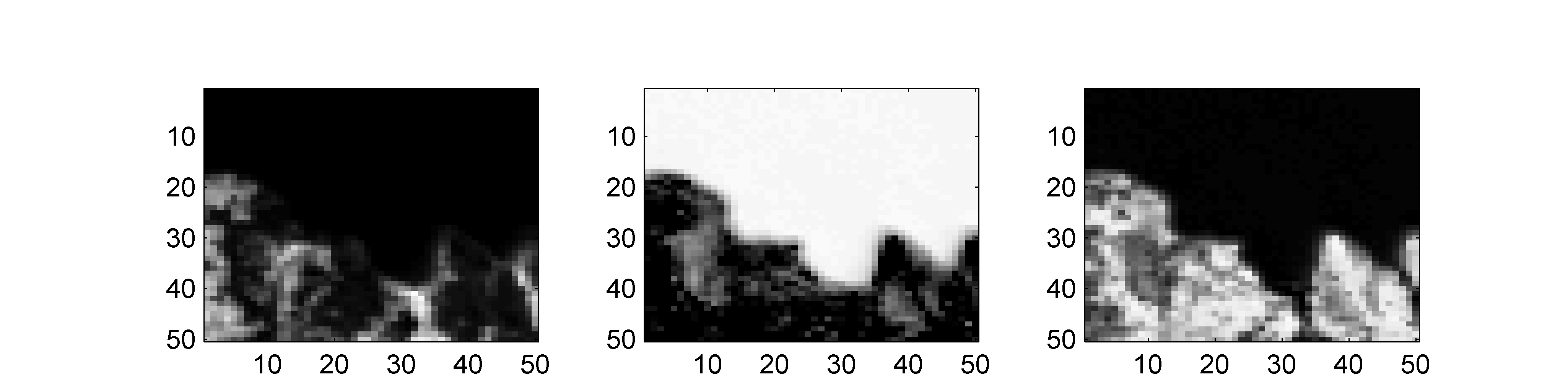}
\caption{Top: abundance maps estimated by the proposed algorithm
(from left to right: vegetation, water and soil). Middle: abundance
maps estimated by the LMM-based Bayesian algorithm (from
    \cite{Dobigeon_IEEE_TSP_2008}). Bottom: fraction maps estimated by the FCLS algorithm
\cite{Heinz2001}.}\label{fig:results_moffett_ab}
\end{figure}

The estimated classification and abundance maps for the proposed
hybrid Gibbs algorithm are depicted in Fig.
\ref{fig:results_moffett_label} (left) and
\ref{fig:results_moffett_ab} (top). The results provided by the
algorithm are very similar and in good agreement with results
obtained on this image with an LMM-based Bayesian algorithm
\cite{Dobigeon_IEEE_TSP_2008} (Fig. \ref{fig:results_moffett_ab},
middle) or with the well-known FCLS algorithm \cite{Heinz2001} (Fig.
\ref{fig:results_moffett_ab}, bottom).

The performance of the proposed algorithm has been also evaluated
for different values of the number of endmembers $R$. The resulting
classification maps for $R=4$ and $R=5$ are given in Fig.
\ref{fig:results_moffett_label} (middle and right). These maps show
that the classification results are quite robust with respect to the
number of endmembers. The corresponding abundance maps can be found
in \cite{Eches2010_techreport_TGRS}, as well as the results of the
proposed algorithm when the number of classes vary.

The computational time of the proposed method (combined with the
N-FINDR procedure) has been compared with the computational times of
two other unmixing algorithms when applied on this Moffett image:
the FCLS algorithm (also combined with N-FINDR) and the constrained
nonnegative matrix factorization (cNMF) algorithm that jointly
estimates the endmember matrix and the abundances \cite{Sajda2004}.
The results\footnote{These simulations have been carried out with an
unoptimized MATLAB $2007$b $32$bit implementation on a
Core(TM)$2$Duo $2.66$GHz computer.} are reported in Table
\ref{tab:time}. The proposed method (referred to as ``Spatial'') has
the higher computational cost when compared to the two others,
mainly due to the joint estimation of the labels and the abundance
vectors. However, it provides more information about unmixing. In
particular, the samples generated with the proposed Gibbs sampler
can be used to determine confidence intervals for the estimated
parameters.

\begin{table}[htbp]
\renewcommand{\arraystretch}{1.2}
\begin{center}
\caption{Computational times of LMM-based unmixing algorithms.}
\begin{tabular}{|c|c|c|c|}
\cline{2-4} \multicolumn{1}{c|}{} & FCLS & cNMF & Spatial  \\
\hline Times (s.)  & $0.388$ & $2.5 \times 10^{3}$ & $8.4 \times 10^{3}$\\
\hline
\end{tabular}
\label{tab:time}
\end{center}
\end{table}

\subsection{Simulation on a larger image}

\begin{figure}[htbp]
\centering
\includegraphics[width=\figwidth]{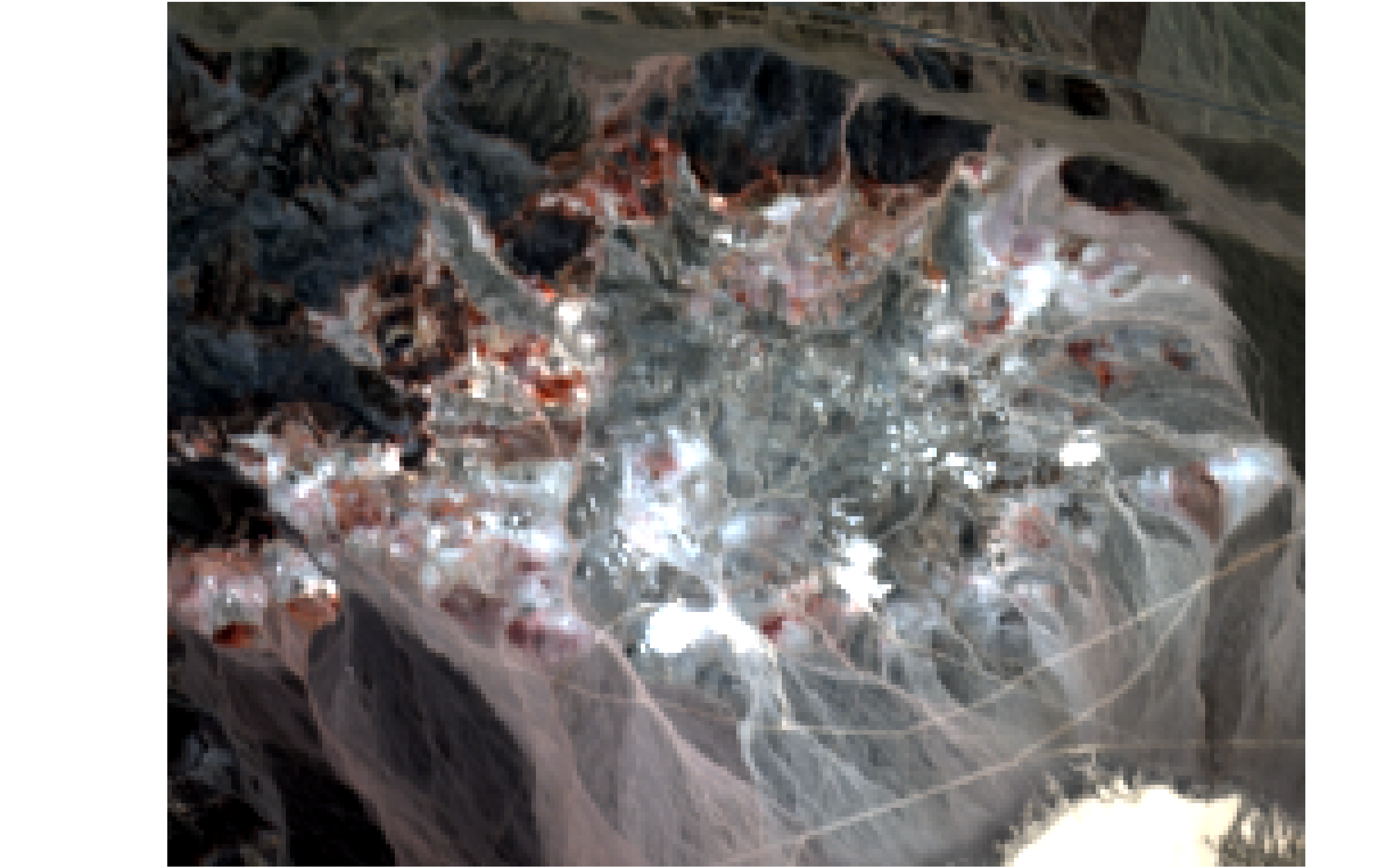}
\caption{AVIRIS image of $190 \times 250$ pixels extracted from
Cuprite scene observed in composite natural
colors.}\label{fig:cuprite_data}
\end{figure}

\begin{figure}[htbp]
\centering
\includegraphics[width=\figwidth]{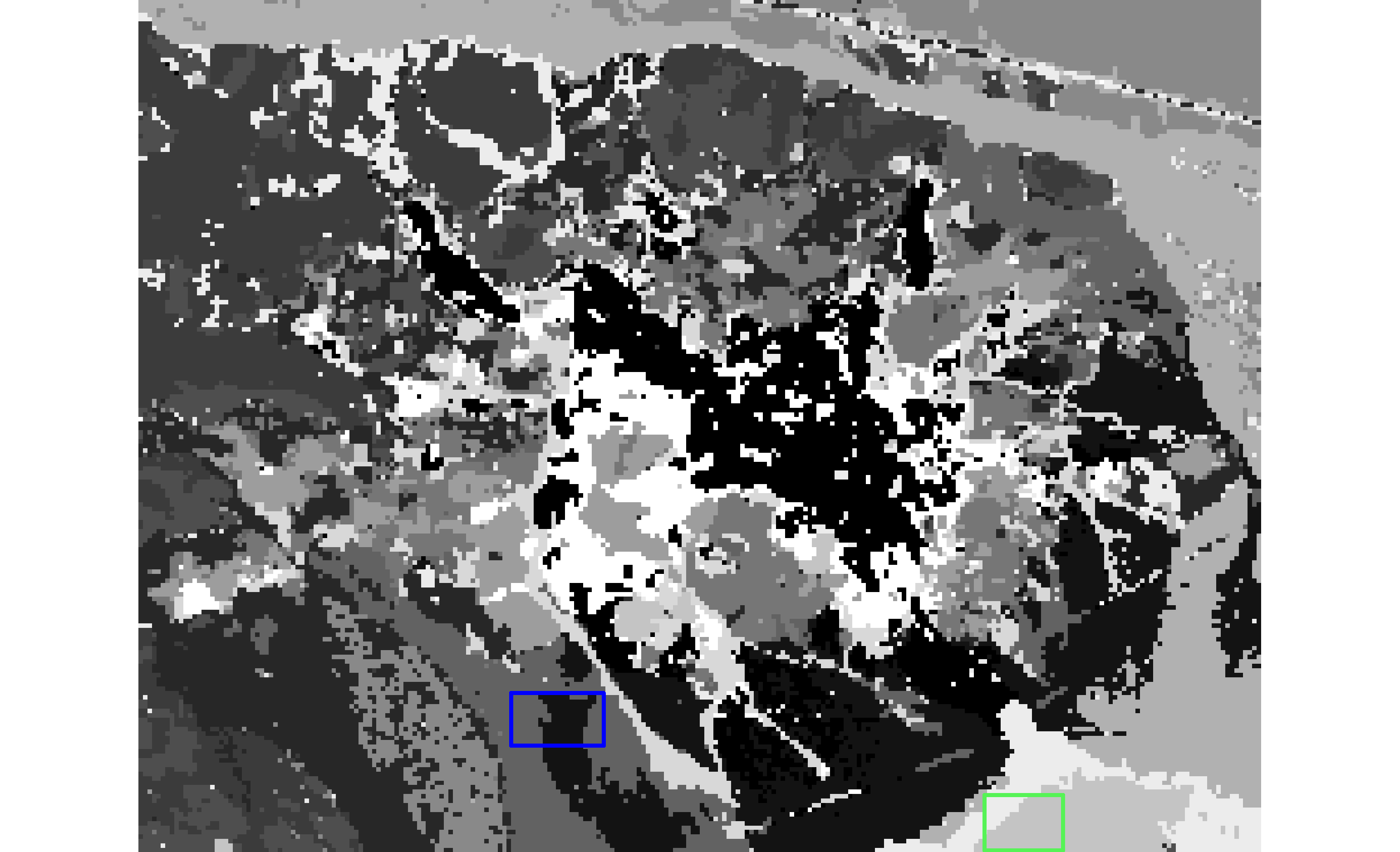}
\caption{Classification map for the $190 \times 250$ Cuprite area
($K=14$).}\label{fig:results_Cuprite_labels}
\end{figure}

\begin{figure*}[htbp]
\includegraphics[width=\figwidthbis]{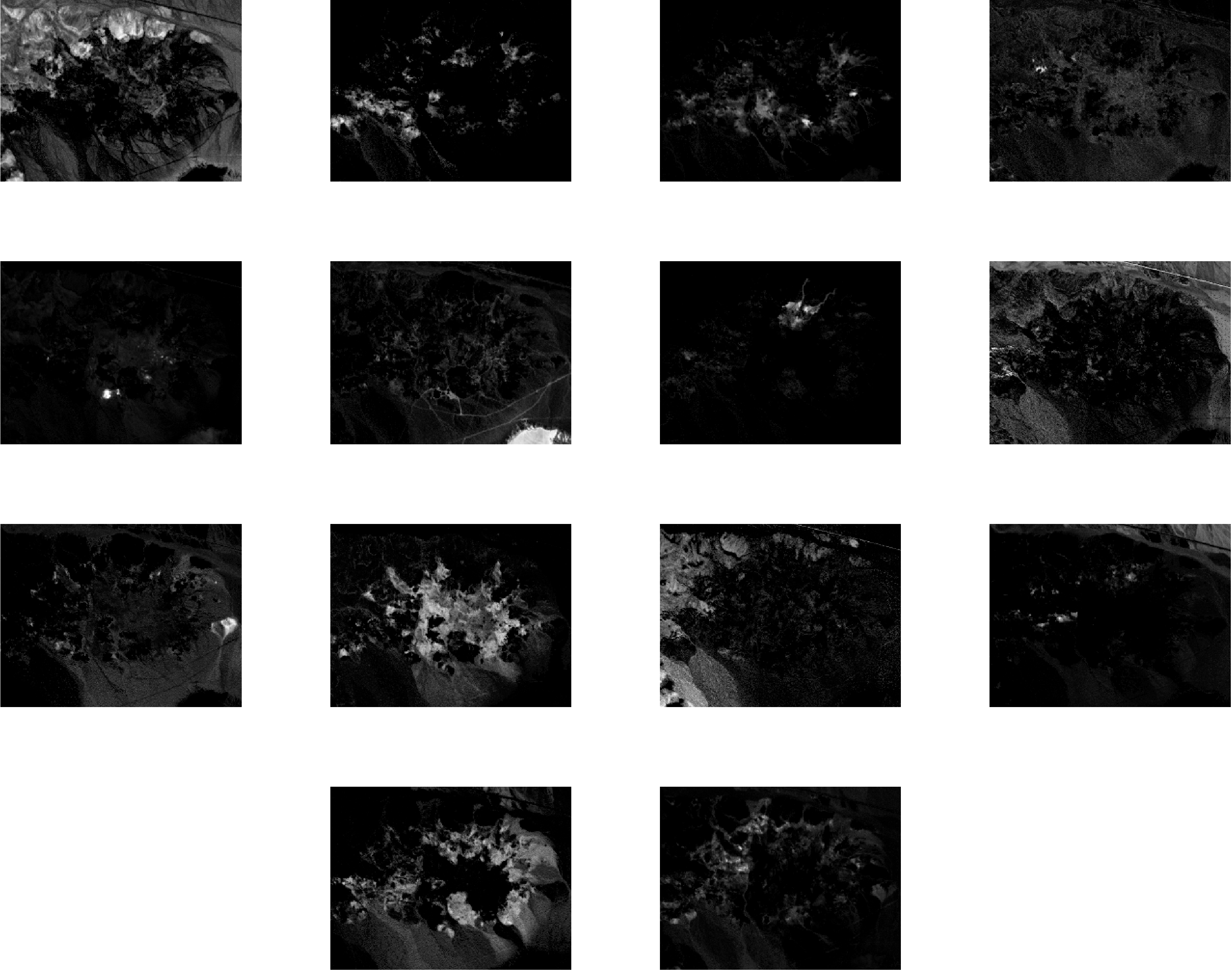}
\caption{Fraction maps of the $190 \times 250$ Cuprite
area.}\label{fig:results_Cuprite_abund}
\end{figure*}

The performance of the proposed Bayesian algorithm has also been
evaluated on a larger real hyperspectral image. The selected scene
has been extracted from the AVIRIS Cuprite image, acquired over a
mining site in Nevada, in $1997$. The geologic characteristics of
the complete data have been mapped in \cite{Clark1993,Clark2003}.
The area of interest of size $190 \times 250$ is represented in Fig.
\ref{fig:cuprite_data} and has been previously studied in
\cite{Nascimento2005} to test the VCA algorithm with $R = 14$.
Therefore, in this experiment, the same number of endmembers has
been extracted by the VCA algorithm. The number of classes has been
set to $K = 14$, which seems to be a sufficient value to capture the
natural diversity of the scene. The proposed algorithm has been used
to estimate the abundance and label maps related to the analyzed
scene. These maps are depicted in Fig.
\ref{fig:results_Cuprite_labels} and
\ref{fig:results_Cuprite_abund}, respectively.

\begin{figure}[htbp]
\centering
\includegraphics[width=\figwidth]{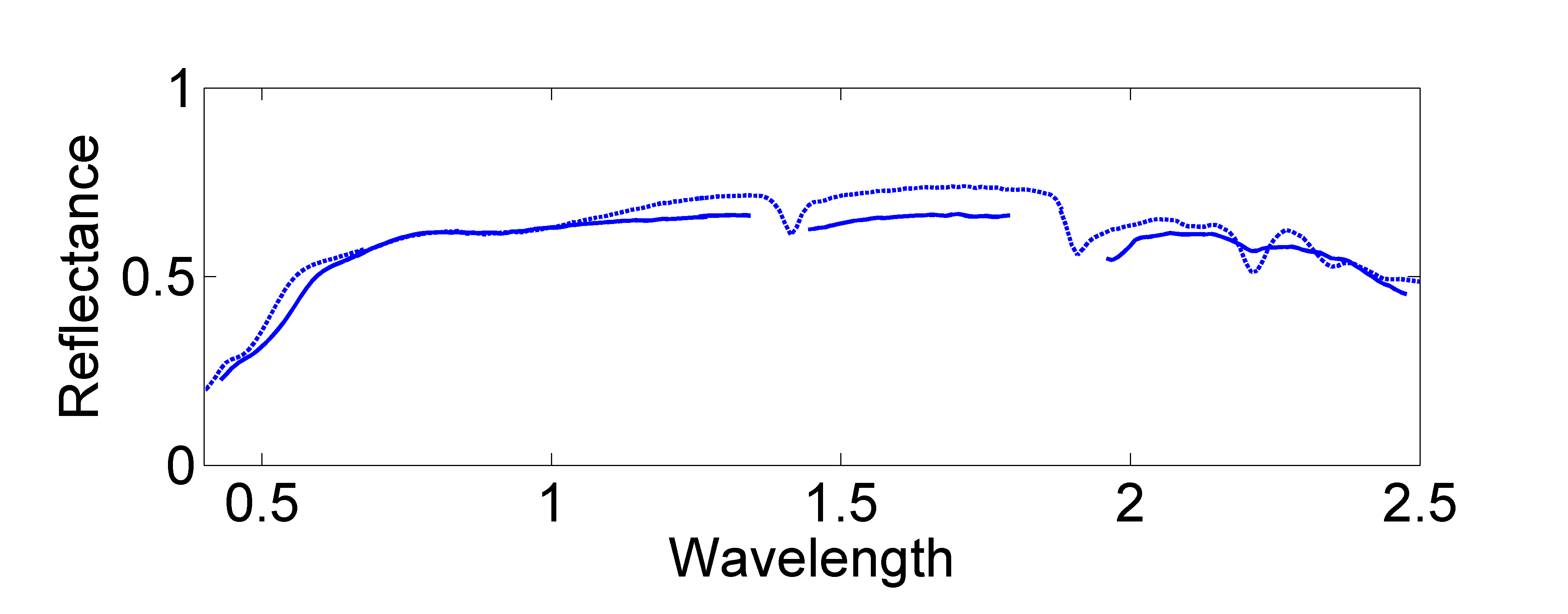}
\caption{Comparison of the $6$th endmember spectrum extracted by the
VCA algorithm (solid line) with the Montmorillonite signature
extracted from the USGS spectral library (dashed
line).}\label{fig:spec_cup_mont}
\end{figure}

The proposed Bayesian inversion algorithm has been able to identify
some regions similar to those recovered in \cite{Nascimento2005}. To
illustrate, the composition of two particular areas (marked as
colored rectangles in Fig.~\ref{fig:results_Cuprite_labels}) is
investigated. Tables \ref{tab:means_cuprite} report the abundance
means for the most significant endmembers that appear in the two
highlighted regions. From these tables, one can conclude that the
two classes represented in black and dark gray of the ``blue'' area
are composed of very mixed pixels (the abundance of the most
significant endmember is $0.201$). On the other hand, both classes
in the ``green'' area are clearly dominated by the $6$th endmember.
By comparing its corresponding signature with the materials included
in the USGS library spectra, this $6$th endmember matches the
Montmorillonite spectrum (see Fig. \ref{fig:spec_cup_mont}). This
result is in good agreement with the ground truth. Indeed, from
\cite{Clark2003}, Montmorillonnite is the most commonly found
material in this area.

\begin{table}[h!]
\renewcommand{\arraystretch}{1.4}
\begin{center}
\caption{Abundance means for the most significant endmembers in each
highlighted region.}

\begin{minipage}[t]{0.4\columnwidth}
    \begin{tabular}{|c|c|c|}
    \cline{2-3}  \multicolumn{1}{c|}{} & \multicolumn{2}{c|}{Green
    area}\\
    \cline{2-3} \multicolumn{1}{c|}{} & light gray  & white \\
    \hline Endm. $1$ & $0.001$ & $0.225$ \\
    \hline Endm. $3$ & $0.045$ & $0.000$ \\
    \hline Endm. $5$ & $0.098$ & $0.027$ \\
    \hline Endm. $6$ & $0.839$ & $0.528$ \\
    \hline
    \end{tabular}
\end{minipage}
\hspace{0.6cm}
\begin{minipage}[t]{0.4\columnwidth}
    \begin{tabular}{|c|c|c|}
    \cline{2-3}  \multicolumn{1}{c|}{} & \multicolumn{2}{c|}{Blue
    area}\\
    \cline{2-3} \multicolumn{1}{c|}{} & black  & dark gray\\
    \hline Endm. $1$ & $0.135$ & $0.044$ \\
    \hline Endm. $9$ & $0.155$ & $0.158$ \\
    \hline Endm. $10$ & $0.159$ & $0.127$ \\
    \hline Endm. $13$ & $0.187$ & $0.206$ \\
    \hline
    \end{tabular}
\end{minipage}
 \label{tab:means_cuprite}
\end{center}
\end{table}

\section{Conclusions}
\label{sec:conclusions} A new hierarchical Bayesian algorithm was
proposed for hyperspectral image unmixing. Markov random fields were
introduced to model spatial correlations between the pixels of the
image. A hidden discrete label was introduced for each pixel of the
image to identify several classes defined by homogeneous abundances
(with constant first and second order statistical moments). The
positivity and sum-to-one constraints on the abundances were handled
by using an appropriate reparametrization defined by logistic
coefficient vectors. We derived the joint posterior distribution of
the unknown parameters and hyperparameters associated to the
proposed Bayesian linear mixing model. An MCMC method was then
studied to generate samples asymptotically distributed according to
this posterior. The generated samples were then used to estimate the
abundance maps as well as the underlying image labels. The results
obtained on simulated data and on real AVIRIS images are very
promising. Future works include the estimation of the granularity
coefficient involved in Potts-Markov random fields.

\section*{Acknowledgments} The
authors would like to thank one of the reviewers for pointing out
the relevant paper \cite{Kent1988} and for his valuable suggestions
that helped to improve the manuscript.

\bibliographystyle{IEEEtran}
\bibliography{reference_olivier}

\begin{thebibliography}{10}
\providecommand{\url}[1]{#1}
\csname url@samestyle\endcsname
\providecommand{\newblock}{\relax}
\providecommand{\bibinfo}[2]{#2}
\providecommand{\BIBentrySTDinterwordspacing}{\spaceskip=0pt\relax}
\providecommand{\BIBentryALTinterwordstretchfactor}{4}
\providecommand{\BIBentryALTinterwordspacing}{\spaceskip=\fontdimen2\font plus
\BIBentryALTinterwordstretchfactor\fontdimen3\font minus
  \fontdimen4\font\relax}
\providecommand{\BIBforeignlanguage}[2]{{%
\expandafter\ifx\csname l@#1\endcsname\relax
\typeout{** WARNING: IEEEtran.bst: No hyphenation pattern has been}%
\typeout{** loaded for the language `#1'. Using the pattern for}%
\typeout{** the default language instead.}%
\else
\language=\csname l@#1\endcsname
\fi
#2}}
\providecommand{\BIBdecl}{\relax}
\BIBdecl

\bibitem{Jusoff2009}
K.~Jusoff, ``Precision forestry using airborne hyperspectral imaging sensor,''
  \emph{Journal of Agricultural Science}, vol.~1, no.~1, pp. 142--147, June
  2009.

\bibitem{Keshava2002}
N.~Keshava and J.~Mustard, ``Spectral unmixing,'' \emph{IEEE Signal Processing
  Magazine}, pp. 44--56, Jan. 2002.

\bibitem{Chang2007}
{C.-I Chang}, \emph{Hyperspectral data exploitation: theory and
  applications}.\hskip 1em plus 0.5em minus 0.4em\relax Hoboken, NJ: Wiley,
  2007.

\bibitem{Martinez2006}
P.~J. Martinez, R.~M. Perez, A.~Plaza, P.~L. Aguilar, M.~C. Cantero, and
  J.~Plaza, ``Endmember extraction algorithms from hyperspectral images,''
  \emph{Annals of Geophysics}, vol.~49, no.~1, pp. 93--101, Feb. 2006.

\bibitem{Heinz2001}
D.~C. Heinz and {C.-I Chang}, ``Fully constrained least squares linear spectral
  mixture analysis method for material quantification in hyperspectral
  imagery,'' \emph{IEEE Trans. Geosci. and Remote Sensing}, vol.~39, no.~3, pp.
  529--545, March 2001.

\bibitem{Theys2009}
C.~Theys, N.~Dobigeon, J.-Y. Tourneret, and H.~Lant\'eri, ``Linear unmixing of
  hyperspectral images using a scaled gradient method,'' in \emph{Proc. IEEE-SP
  Workshop Stat. and Signal Processing (SSP)}, Cardiff, UK, Aug. 2009, pp.
  729--732.

\bibitem{Dobigeon_IEEE_TSP_2008}
N.~Dobigeon, J.-Y. Tourneret, and {{C.-I} Chang}, ``Semi-supervised linear
  spectral using a hierarchical {B}ayesian model for hyperspectral imagery,''
  \emph{IEEE Trans. Signal Processing}, vol.~56, no.~7, pp. 2684--2696, July
  2008.

\bibitem{Besag1974}
J.~Besag, ``Spatial interaction and the statistical analysis of lattice
  systems,'' \emph{J. Roy. Stat. Soc. Ser. B}, vol.~36, no.~2, pp. 192--236,
  1974.

\bibitem{Geman1984}
S.~Geman and D.~Geman, ``Stochastic relaxation, {G}ibbs distributions, and the
  {B}ayesian restoration of images,'' \emph{IEEE Trans. Patt. Anal. Mach.
  Intell.}, vol.~6, no.~6, pp. 721--741, Nov. 1984.

\bibitem{Kent1988}
J.~T. Kent and K.~V. Mardia, ``Spatial classification using fuzzy membership
  models,'' \emph{IEEE Trans. Patt. Anal. Mach. Intell.}, vol.~10, no.~5, pp.
  659--671, 1988.

\bibitem{Dobigeon_IEEE_TSP_2009}
N.~Dobigeon, S.~Moussaoui, M.~Coulon, J.-Y. Tourneret, and A.~O. Hero, ``Joint
  {B}ayesian endmember extraction and linear unmixing for hyperspectral
  imagery,'' \emph{IEEE Trans. Signal Processing}, vol.~57, no.~11, pp.
  4355--4368, Nov. 2009.

\bibitem{Wu1982}
F.~Wu, ``The {P}otts model,'' \emph{Rev. Modern Phys.}, vol.~54, no.~1, pp.
  235--268, Jan. 1982.

\bibitem{Mohammadpour2004}
A.~{Mohammadpour}, O.~{F{\'e}ron}, and A.~{Mohammad-Djafari}, ``Bayesian
  segmentation of hyperspectral images,'' in \emph{Proc. Workshop on Bayesian
  Inference and Maximum Entropy Methods (MaxEnt)}, ser. AIP Conference Series,
  R.~Fischer, R.~Preuss, and U.~V. Toussaint, Eds., vol. 735, Max Plank Inst.,
  Germany, Nov. 2004, pp. 541--548.

\bibitem{Rellier2004}
G.~Rellier, X.~Descombes, F.~Falzon, and J.~Zerubia, ``Texture feature analysis
  using a {G}auss-{M}arkov model in hyperspectral image classification,''
  \emph{IEEE Trans. Geosci. and Remote Sensing}, vol.~42, no.~7, pp.
  1543--1551, July 2004.

\bibitem{Neher2005}
R.~Neher and A.~Srivastava, ``A {B}ayesian {M}{R}{F} framework for labeling
  terrain using hyperspectral imaging,'' \emph{IEEE Trans. Geosci. and Remote
  Sensing}, vol.~43, no.~6, pp. 1363--1374, June 2005.

\bibitem{Feron2005}
O.~Féron and A.~Mohammad-Djafari, ``Image fusion and unsupervised joint
  segmentation using a {HMM} and {MCMC} algorithms,'' \emph{J. Electron.
  Imaging}, vol.~14, no.~2, May 2005.

\bibitem{Bali2008}
N.~Bali and A.~Mohammad-Djafari, ``Bayesian approach with hidden {M}arkov
  modeling and mean field approximation for hyperspectral data analysis,''
  \emph{IEEE Trans. Image Processing}, vol.~17, no.~2, pp. 217--225, Feb. 2008.

\bibitem{Li2010}
J.~Li, J.~M. Bioucas-Dias, and A.~Plaza, ``Semi-supervised hyperspectral image
  segmentation using multinomial logistic regression with active learning,''
  \emph{IEEE Trans. Geosci. and Remote Sensing}, vol.~48, no.~11, pp.
  4085--4098, Nov. 2010.

\bibitem{Fauvel2008}
M.~Fauvel, J.~A. Benediktsson, J.~Chanussot, and J.~R. Sveinsson, ``Spectral
  and spatial classification of hyperspectral data using {S}{V}{M}s and
  morphological profiles,'' \emph{IEEE Trans. Geosci. and Remote Sensing},
  vol.~46, no.~11, pp. 3804--3814, Nov. 2008.

\bibitem{Tarabalka2010}
Y.~Tarabalka, M.~Fauvel, J.~Chanussot, and J.~A. Benediktsson, ``{SVM} and
  {MRF}-based method for accurate classification of hyperspectral images,''
  \emph{IEEE Geosci. and Remote Sensing Letters}, vol.~7, no.~4, pp. 736--740,
  Oct. 2010.

\bibitem{Eches_IEEE_WHISPERS_2009}
O.~Eches, N.~Dobigeon, and J.-Y. Tourneret, ``{NCM}-based {B}ayesian algorithm
  for hyperspectral unmixing,'' in \emph{Proc. IEEE GRSS Workshop on
  Hyperspectral Image and SIgnal Processing: Evolution in Remote Sensing
  (WHISPERS)}, Grenoble, France, Aug. 2009, pp. 1--4.

\bibitem{Harsanyi1994}
J.~Harsanyi and {C.-I Chang}, ``Hyperspectral image classification and
  dimensionality reduction: An orthogonal subspace projection approach,''
  \emph{IEEE Trans. Geosci. and Remote Sensing}, vol.~32, pp. 779--785, July
  1994.

\bibitem{Chang1998}
{C.-I Chang}, ``Further results on relationship between spectral unmixing and
  subspace projection,'' \emph{IEEE Trans. Geosci. and Remote Sensing},
  vol.~36, no.~3, pp. 1030--1032, May 1998.

\bibitem{Chang1998b}
{C.-I Chang}, X.-L. Zhao, M.~L.~G. Althouse, and J.~J. Pan, ``Least squares
  subspace projection approach to mixed pixel classification for hyperspectral
  images,'' \emph{IEEE Trans. Geosci. and Remote Sensing}, vol.~36, no.~3, pp.
  898--912, May 1998.

\bibitem{Kevrann1995}
C.~Kevrann and F.~Heitz, ``A {M}arkov random field model-based approach to
  unsupervised texture segmentation using local and global statistics,''
  \emph{IEEE Trans. Image Processing}, vol.~4, no.~6, pp. 856--862, June 1995.

\bibitem{Tonazzini2006}
A.~Tonazzini, L.~Bedini, and E.~Salerno, ``A {M}arkov model for blind image
  separation by a mean-field {E}{M} algorithm,'' \emph{IEEE Trans. Image
  Processing}, vol.~15, no.~2, pp. 473--481, Feb. 2006.

\bibitem{Rand2003}
R.~S. Rand and D.~M. Keenan, ``Spatially smooth partitioning of hyperspectral
  imagery using spectral/spatial measures of disparity,'' \emph{IEEE Trans.
  Geosci. and Remote Sensing}, vol.~41, no.~6, pp. 1479--1490, June 2003.

\bibitem{Kindermann1980}
R.~Kindermann and J.~L. Snell, \emph{Markov random fields and their
  applications}.\hskip 1em plus 0.5em minus 0.4em\relax Providence, RI: Amer.
  Math. Soc., 1980.

\bibitem{Eches2010_techreport_TGRS}
\BIBentryALTinterwordspacing
O.~Eches, N.~Dobigeon, and J.-Y. Tourneret, ``Enhancing hyperspectral image
  unmixing with spatial correlations,'' University of Toulouse, Tech. Rep.,
  July 2010. [Online]. Available: \url{http://eches.perso.enseeiht.fr}
\BIBentrySTDinterwordspacing

\bibitem{Marin2007}
J.-M. Marin and C.~P. Robert, \emph{Bayesian core: a practical approach to
  computational {B}ayesian statistics}.\hskip 1em plus 0.5em minus 0.4em\relax
  New-York: Springer, 2007.

\bibitem{Ripley1988}
B.~D. Ripley, \emph{Statistical inference for spatial processes}.\hskip 1em
  plus 0.5em minus 0.4em\relax Cambridge: Cambridge University Press, 1988.

\bibitem{Zhou1997}
Z.~Zhou, R.~Leahy, and J.~Qi, ``Approximate maximum likelihood hyperparameter
  estimation for {G}ibbs prior,'' \emph{IEEE Trans. Image Processing}, vol.~6,
  no.~6, pp. 844--861, June 1997.

\bibitem{Descombes1999}
X.~Descombes, R.~Morris, J.~Zerubia, and M.~Berthod, ``Estimation of {M}arkov
  random field prior parameters using {M}arkov chain {M}onte {C}arlo maximum
  likelihood,'' \emph{IEEE Trans. Image Processing}, vol.~8, no.~7, pp.
  945--963, July 1999.

\bibitem{Celeux2003}
G.~Celeux, F.~Forbes, and N.~Peyrard, ``{E}{M} procedures using mean field-like
  approximations for {M}arkov model-based image segmentation,'' \emph{Pattern
  Recognition}, vol.~36, no.~1, pp. 131--144, 2003.

\bibitem{Gelman1996}
A.~Gelman, F.~Bois, and J.~Jiang, ``Physiological pharmacokinetic analysis
  using population modeling and informative prior distributions,'' \emph{J.
  Amer. Math. Soc.}, vol.~91, no. 436, pp. 1400--1412, Dec. 1996.

\bibitem{Themelis2008}
K.~Themelis and A.~A. Rontogiannis, ``A soft constrained {M}{A}{P} estimator
  for supervised hyperspectral signal unmixing,'' in \emph{Proc. of European
  Signal Processing Conf. (EUSIPCO)}, Lausanne, Switzerland, Aug. 2008.

\bibitem{Punskaya2002}
E.~Punskaya, C.~Andrieu, A.~Doucet, and W.~Fitzgerald, ``{B}ayesian curve
  fitting using {MCMC} with applications to signal segmentation,'' \emph{IEEE
  Trans. Signal Processing}, vol.~50, no.~3, pp. 747--758, March 2002.

\bibitem{Tourneret2007}
N.~Dobigeon, J.-Y. Tourneret, and M.~Davy, ``Joint segmentation of piecewise
  constant autoregressive processes by using a hierarchical model and a
  {B}ayesian sampling approach,'' \emph{IEEE Trans. Signal Processing},
  vol.~55, no.~4, pp. 1251--1263, April 2007.

\bibitem{Robert2007choice}
C.~P. Robert, \emph{The Bayesian Choice: from Decision-Theoretic Motivations to
  Computational Implementation}, 2nd~ed., ser. Springer Texts in
  Statistics.\hskip 1em plus 0.5em minus 0.4em\relax New York: Springer-Verlag,
  2007.

\bibitem{Dobigeon_IEEE_TSP_2007b}
N.~Dobigeon, J.-Y. Tourneret, and M.~Davy, ``Joint segmentation of piecewise
  constant autoregressive processes by using a hierarchical model and a
  {B}ayesian sampling approach,'' \emph{IEEE Trans. Signal Processing},
  vol.~55, no.~4, pp. 1251--1263, April 2007.

\bibitem{Robert2004}
C.~P. Robert and G.~Casella, \emph{{M}onte {C}arlo Statistical Methods},
  2nd~ed.\hskip 1em plus 0.5em minus 0.4em\relax New York: Springer Verlag,
  2004.

\bibitem{Roberts1996}
G.~O. Roberts, ``Markov chain concepts related to sampling algorithms,'' in
  \emph{{M}arkov Chain {M}onte {C}arlo in Practice}, W.~R. Gilks,
  S.~Richardson, and D.~J. Spiegelhalter, Eds.\hskip 1em plus 0.5em minus
  0.4em\relax London: Chapman \& Hall, 1996, pp. 259--273.

\bibitem{ENVImanual2003}
{{RSI} (Research Systems Inc.)}, \emph{ENVI User's guide Version 4.0}, Boulder,
  CO 80301 USA, Sept. 2003.

\bibitem{Christophe2005}
E.~Christophe, D.~L\'eger, and C.~Mailhes, ``Quality criteria benchmark for
  hyperspectral imagery,'' \emph{IEEE Trans. Geosci. and Remote Sensing},
  vol.~43, no.~9, pp. 2103--2114, Sept. 2005.

\bibitem{Akgun2005}
T.~Akgun, Y.~Altunbasak, and R.~M. Mersereau, ``Super-resolution reconstruction
  of hyperspectral images,'' \emph{IEEE Trans. Image Processing}, vol.~14,
  no.~11, pp. 1860--1875, Nov. 2005.

\bibitem{Winter1999}
M.~E. Winter, ``Fast autonomous spectral endmember determination in
  hyperspectral data,'' in \emph{Proc. 13th Int. Conf. on Applied Geologic
  Remote Sensing}, vol.~2, Vancouver, April 1999, pp. 337--344.

\bibitem{Sajda2004}
P.~Sajda, S.~Du, T.~R. Brown, R.~Stoyanova, D.~C. Shungu, X.~Mao, and L.~C.
  Parra, ``Nonnegative matrix factorization for rapid recovery of constituent
  spectra in magnetic resonance chemical shift imaging of the brain,''
  \emph{IEEE Trans. Medical Imaging}, vol.~23, no.~12, pp. 1453–--1465, 2004.

\bibitem{Clark1993}
R.~N. Clark, G.~A. Swayze, and A.~Gallagher, ``Mapping minerals with imaging
  spectroscopy, {U.S. Geological Survey},'' \emph{Office of Mineral Resources
  Bulletin}, vol. 2039, pp. 141--150, 1993.

\bibitem{Clark2003}
R.~N. {{Clark \textit{et al.}}}, ``Imaging spectroscopy: {E}arth and planetary
  remote sensing with the {USGS Tetracorder} and expert systems,'' \emph{J.
  Geophys. Res.}, vol. 108, no. E12, pp. 5--1--5--44, Dec. 2003.

\bibitem{Nascimento2005}
J.~M. Nascimento and J.~M. {Bioucas-Dias}, ``Vertex component analysis: a fast
  algorithm to unmix hyperspectral data,'' \emph{IEEE Trans. Geosci. and Remote
  Sensing}, vol.~43, no.~4, pp. 898--910, April 2005.

\end{thebibliography}
\end{document}